\documentclass{article}
\usepackage[utf8]{inputenc}
\usepackage[font={footnotesize}]{caption}
\usepackage[margin=1.in]{geometry}
\usepackage{array}
\usepackage{graphicx}
\usepackage{hyperref}

\usepackage{amsmath,amssymb,mathrsfs}
\usepackage{tabularx,booktabs,multirow}
\usepackage{color,colortbl}
\usepackage[dvipsnames]{xcolor}
\usepackage{authblk}

\definecolor{pierred}{rgb}{0.9, 0.1, 0.1}

\definecolor{cristianoblue}{rgb}{0.1, 0.1, 0.8}

\definecolor{elenaorange}{rgb}{1,0.5,0}

\title{NREM and REM: cognitive and energetic gains in thalamo-cortical sleeping and awake spiking model}

\author [1, 2, *] {Chiara De Luca}
\author [1, *] {Leonardo Tonielli}
\author [1] {Elena Pastorelli}
\author [1] {Cristiano Capone}
\author [1] {Francesco Simula}
\author [1] {Cosimo Lupo}
\author [1] {Irene Bernava}
\author [3,4] {Gianmarco Tiddia}
\author [1] {Giulia De Bonis}
\author [3,4] {Bruno Golosio}
\author [1] {Pier Stanislao Paolucci}

\affil [1] {Istituto Nazionale di Fisica Nucleare, Sezione di Roma}
\affil [2] {PhD Program in Behavioural Neuroscience, ``Sapienza'' University of Rome}
\affil [3] {Dipartimento di Fisica, Università di Cagliari, Italy}
\affil [4] {Istituto Nazionale di Fisica Nucleare, Sezione di Cagliari, Italy}
\affil [*] {These authors contributed equally to this work}

\date{\today}

\begin{document}

\maketitle

\begin{abstract}
Sleep is essential for learning and cognition, but the mechanisms by which it stabilizes learning, supports creativity, and manages the energy consumption of networks engaged in post-sleep task have not been yet modelled. During sleep, the brain cycles between non-rapid eye movement (NREM), a mainly unconscious state characterized by collective oscillations, and rapid eye movement (REM), associated with the integrated experience of dreaming. We propose a biologically grounded two-area thalamo-cortical plastic spiking neural network model and investigate the role of NREM - REM cycles on its awake performance. We demonstrate that sleep has a positive effect on energy consumption and cognitive performance during the post-sleep awake classification task of handwritten digits. NREM and REM simulated dynamics modify the synaptic structure into a sharper representation of training experiences. Sleep-induced synaptic modifications reduce firing rates and synaptic activity without reducing cognitive performance. Also, it creates novel multi-area associations. The model leverages the apical amplification, isolation and drive experimentally grounded principles and the combination of contextual and perceptual information. In summary, the main novelty is the proposal of a multi-area plastic model that also expresses REM and integrates information during a plastic dream-like state, with cognitive and energetic benefits during post-sleep awake classification.
\end{abstract}

\section{Introduction}
\label{sec:intro}

During wakefulness, the perceptual system is continuously subjected to sensory input from different sources and modalities. The involved brain areas process the current input in a framework set by previous knowledge (acquired through individual and evolutionary experience), with a crucial role played by the exchange of signals with other brain areas. The ability of the brain to integrate and segregate this information by building a subjective, though coherent and complete representation of the environment -- involving cognitive processes such as learning, decision-making, and selective attention~\cite{Mejer2019} -- is impressive.

Although there is plenty of empirical evidences suggesting that the nervous system uses a statistically optimal approach in combining external information from previous experiences with the current one, the understanding of how the brain implements these strategies is still a topic of research. For example, the investigation of multi-sensory processing at the neuronal level in terms of multiple-cue integration, instead of modality-specific (or stand-alone) single-cue integration, provides a more reliable estimate of e.g. objects and events, also leading to behavioral benefits such as faster and more accurate responses to given situations~\cite{Gingras2009}.

In this framework, sleep plays a central role. Indeed, it is known to be essential in all animal species, especially in younger individuals, for their brain circuit development~\cite{Frank2001}. Further profound positive effects of the sleep on the body in general and the brain in particular notably involve the regulation of the endocrine and immune systems~\cite{Besedovsky2012} and the extracellular clearance of potentially toxic substances~\cite{xie2013}. Finally, among the mechanisms occurring during the sleep and yet to be investigated, there is its central role in the recovery from brain damages~\cite{Siccoli2008}. On the other side, instead, sleep deprivation is well known to have detrimental effects on cognition~\cite{Killgore2010}.

More in detail, among the cognitive benefits of sleep we are most interested in here, we have the consolidation of learned information, the creation of novel associations, and the preparation for tasks expected during the next awake periods~\cite{Buzsaki2015, Sejnowski2000}, optimizing the related post-sleep energy consumption. In this regard, evidences are mounting in support of the Synaptic Homeostasis Hypothesis (SHY) model proposed in~\cite{Tononi2014}, suggesting the crucial role of sleep in facilitating the recovery of neuronal energy and synaptic resources that have been depleted after prolonged waking~\cite{Tononi2014, tononi2019}. This mechanism implies the normalization of memory representation and the optimization of energetic working point of the system by recalibrating synaptic weights~\cite{Tononi2014} and firing rates~\cite{watson2016}. Additionally, the consolidation of memory during sleep can produce a strengthening of associations, together with qualitative changes also in the memory representation, leading to an improved resistance to interferences from another similar tasks~\cite{ellenbogen2006, korman2007} and improved performances at re-testing, in absence of additional practice during the retention interval~\cite{walker2003, tucker2006}.

Sleep is not characterized by homogeneous features, rather it exhibits unique patterns of cortical activity alternating between Non-Rapid-Eye-Movement (NREM) stages, with a predominant high-amplitude, low-frequency (``slow wave'') EEG activity in the delta band ($0.5-4$Hz) and Rapid-Eye-Movement (REM) stages, with an activity mainly shifted to the theta band ($4-11$Hz)~\cite{brown2012}. The alternance of NREM and REM phases during the normal sleep cycle suggests that they have complementary roles, in particular for what concerns memory optimization. During NREM, indeed, system consolidation is promoted through the reactivation and the redistribution of selected memory traces for long-term storage, while REM sleep might help to stabilize transformed memories by enabling undisturbed synaptic consolidation~\cite{diekelmann2010}.

NREM and REM sleep phases also differ for the levels of consciousness they exhibit, and both clearly differ from wakefulness, too. The cortical activity during these three stages has been studied with different techniques~\cite{watson2016, niethard2016}, and in particular their underlying mechanisms of consciousness have been studied through Transcranial Magnetic Stimulation together with high-density Electroencephalography (TMS/EEG), measuring the transmission mechanism following the stimulation of a single cortical area to the rest of the brain~\cite{massimini2005, sarasso2014, pigorini2015}. From the theoretical point of view, consciousness is believed to require the joint presence of functional integration and functional differentiation, otherwise defined as \textit{brain complexity}; these studies invariably show that the complexity of the cortical response to TMS/EEG collapses when consciousness is lost during deep sleep, anesthesia, and vegetative states following severe brain injuries, while it recovers when consciousness resurges in spontaneous wakefulness, during dreaming, and in the minimally conscious state or ``locked-in'' syndrome~\cite{sarasso2014}. The fading of consciousness during certain stages of sleep may be related to a breakdown in effective cortical connectivity~\cite{sarasso2014}; there are evidences in this direction for trans-callosal and long-range connectivity during NREM sleep, leading to the inability for the cortical areas to effectively interact, in contrast to the persistence (or the increase) in inter-hemispheric and inter-areal broadband coherence~\cite{kandel2000}. In this stage, cortical activations become more local and stereotypical, indicating a significant impairment of the intra-cortical dialogue. Dream-like consciousness, on the other hand, occurs during various (mainly REM) phases of sleep, also including sleep onset and late night. In these stages, TMS/EEG triggers more widespread and differentiated patterns of cortical activation, similar to those observed in wakefulness~\cite{massimini2010}.

However, the neuronal mechanisms underlying dreaming experiences during the sleep remain an open question. Several hypotheses and theories have been proposed in this regard, trying to explain the mechanisms and functions of dreaming~\cite{nir2010}. In this paper, we rely on a specific cellular mechanism proposed in~\cite{aru2020}, assuming a top-down explanation for dreams: like imagination, they start from abstract thoughts, concepts, or even unconscious wishes, and are secondarily enriched with sensory precepts. The three main brain states (wakefulness, REM/dream sleep and NREM/dreamless sleep) can be then defined by relying on the different roles played by acetylcholine (ACh) and noradrenaline (NA) neuro-modulators. ACh regulates the transmission of information from the apical integration zone to the soma of neurons, while NA regulates the extent of spatio-temporal summation of sensorial input to the apical dendrites. During quiet and active wakefulness, high levels of the two neuro-modulators facilitate the transmission of relevant information (\textit{apical amplification}). During dreams, at variance, lower levels of NA lead to a predominance of apical internal input, letting contextual hints to turn into self-fulfilling prophecies (\textit{apical drive}). Finally, during dreamless sleep, both NA and ACh levels are low, so that neurons do not integrate neither apical nor sensorial inputs any more (\textit{apical isolation}).
However, more complex states on the whole brain scale can be observed, as combinations of the three defined above: e.g., some areas entering into NREM- or REM-like states, with others still awake and active, or the coexistence of NREM and REM sleep phases in different brain areas (see e.g.~\cite{2021BrainRev-Frohlich-ConsiousnessDeltaParadox}). In this work, we will not consider such combinations of different local states.

Building on the concepts exposed above, in this paper we present a plastic spiking model capable of expressing awake-, NREM- and REM-like dynamics and matching several experimental observations~\cite{watson2016}. The model expands the data-inspired Thalamo-Cortical (ThaCo) plastic spiking network models already proposed in~\cite{capone2019, golosio2021}, which were able to access only the awake and NREM sleep-like states. There, the combination of context and perception in a single-area thalamo-cortical model has been exploited, relying on a soft winner-take-all circuit of excitatory and inhibitory spiking neurons, calibrating the circuit to express awake and deep-sleep states with features comparable with biological measures. The capacity to incrementally learn from a few examples, its resilience when proposed with noisy perceptions and contextual signals, and an improvement in visual classification after NREM sleep due to induced synaptic homeostasis and association of similar memories have also been demonstrated in~\cite{capone2019, golosio2021}. In this work, we pass to a two-area model, mimicking the two hemifields by getting as input two different portions of the external stimulus. The input preprocessing has also been improved, implemented in a foveal-like protocol.

\section{Results}
\label{sec:results}

In this paper, our aim is to model and implement a biologically-plausible thalamo-cortical network capable of expressing awake, NREM, and REM states with features comparable to experimental knowledge and theoretical principles. In continuity with the model presented in~\cite{golosio2021}, we rely mainly on biological observations made in~\cite{watson2016} on the changes of firing rate distributions in awake, sleep, and post-sleep phases, and in~\cite{gonzalez2019} for what concerns power spectral density measurements in different brain states.
Here, we show that this model is capable of performing multiple awake-NREM-REM cycles, proving the beneficial cognitive and energetic effects of such dynamics on both classification performances and network internal structure.

\begin{figure}[!th]
    \centering
    \includegraphics[width=0.8\textwidth]{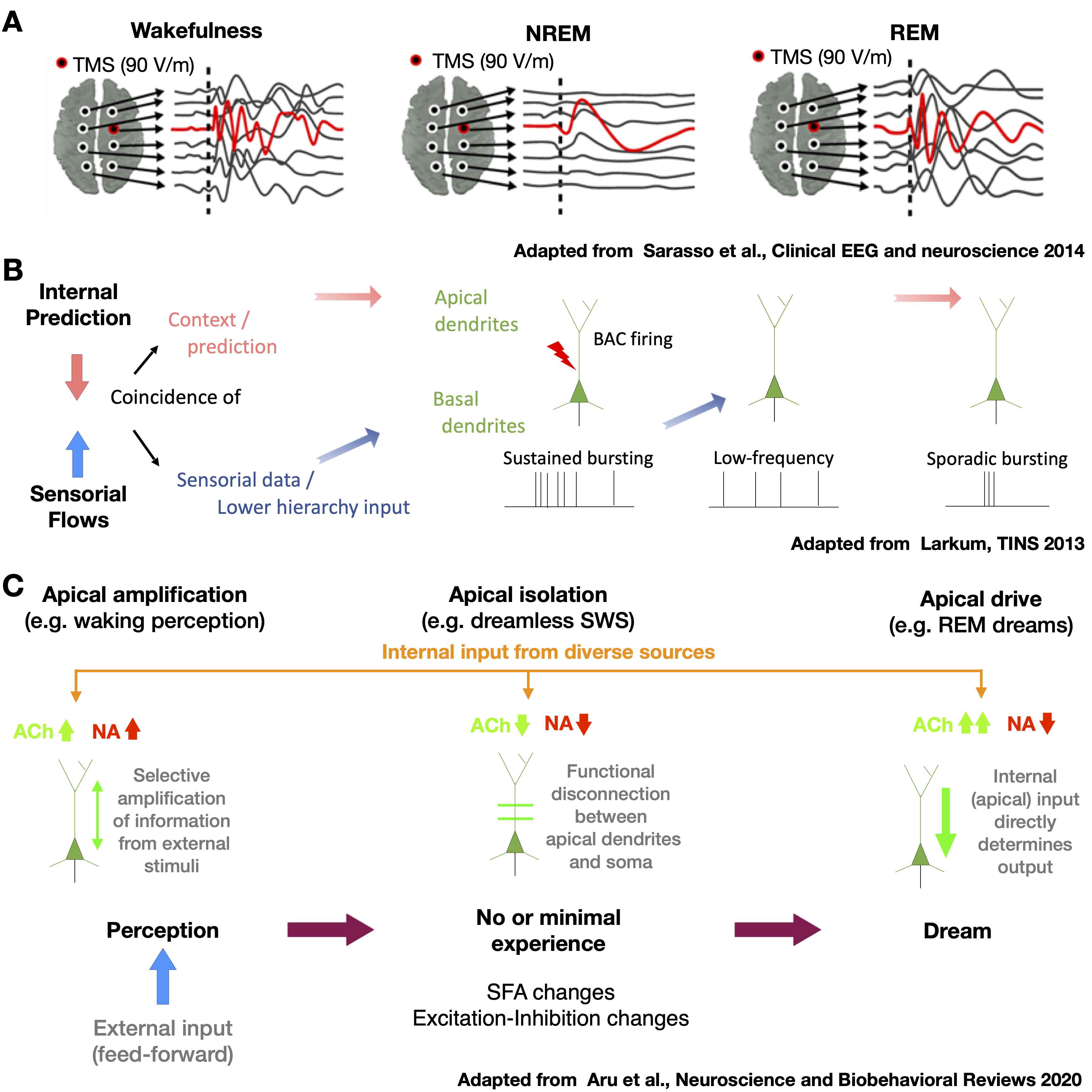}
    \caption{
    \textbf{Ingredients to build the network model.}
    \textbf{A}) Figure adapted from~\cite{sarasso2014}, depicting the TMS/EEG response in different brain states. Wakefulness is associated with spatially and temporally differentiated patterns of activation; NREM sleep features a loss in the ability to engage in complex activity patterns (loss of integration and differentiation/information); REM sleep shows a recovery of recurrent waves of activity associated with spatially and temporally differentiated patterns of activation.
    \textbf{B}) Figure adapted from~\cite{aruLarkum2020}, depicting a conceptual representation of the Back-propagation Activated Calcium (BAC) firing hypothesis. Pyramidal neurons receiving predominantly feed-forward information are likely to fire steadily at low rates, whereas the simultaneous presence of contextual and perceptual streams changes the mode of firing into bursts (BAC firing).
    \textbf{C}) Figure adapted from~\cite{aru2020}, considering the role of acetylcholine (ACh) and noradrenaline (NA) neuro-modulators during wakefulness and NREM and REM sleep phases, introducing the concepts of apical amplification, apical isolation, and apical drive.
    }
    \label{fig:exp_ground}
\end{figure}

\subsection{The model}
\label{res:subsec:model}
\begin{figure}[!th]
    \centering
    \includegraphics[width=0.8\textwidth]{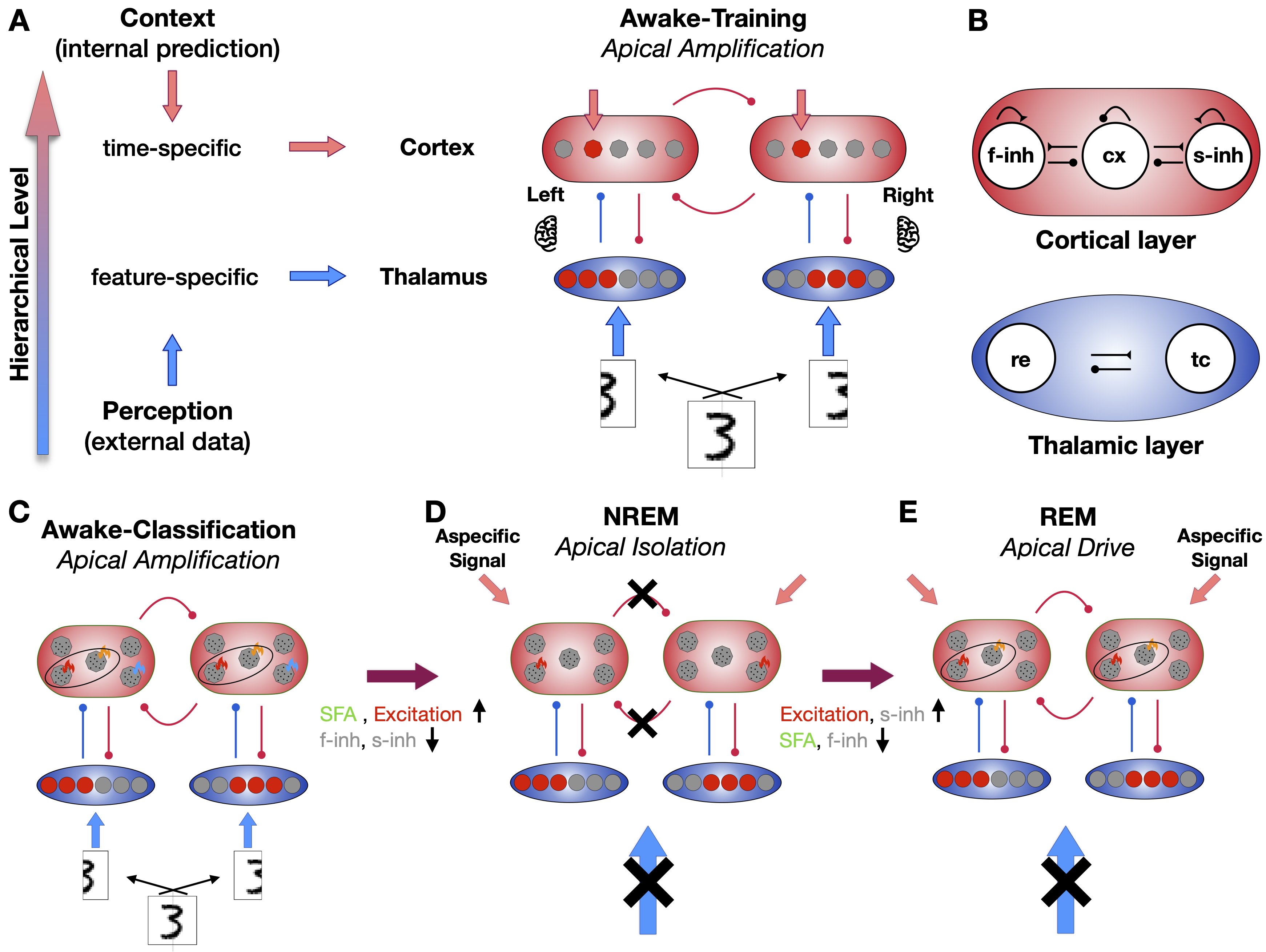}
    \caption{
    \textbf{Sketch of the structure of the two-area thalamo-cortical model and protocol description.}
    \textbf{A}) Two layers are implemented: thalamic and cortical. Awake learning stage mimicking the \textit{Apical Amplification} mechanism: different portion of the perceptual feature-specific input is fed independently towards the thalamic layer in each area. Also a lateral contextual stimulus is provided to a time-specific subset of neurons in the cortical layer (\textit{cx}), mimicking information coming from other higher-level areas in the brain. Due to temporal coincidence of sensory and contextual stimuli, specific neural groups are selected to encode perceptual information in order to create a multi-area cortical representation of the percept. Inhibitory network manage an hard winner-takes-all dynamics (hard-WTA) (i.e. only the chosen groups are permitted to activate, see Sec.~\ref{subsec:methods:brain_states} for more details))
    \textbf{B}) Internal composition of the implemented layers: the thalamic layer composed of an excitatory and an inhibitory population (\textit{tc}, \textit{re}); cortical layer composed of an excitatory population (\textit{cx}) connected to a fast-spiking inhibitory population (\textit{f-in}) through strong connections and a slow-spiking inhibitory population (\textit{s-in}) through weak synapses. The two \textit{cx} population in each area are reciprocally connected.  
    \textbf{C}) Awake classification stage: perceptual stimulus only is fed into the network. All connections are active and synaptic STDP is set off. The coloured flames aim to depict different levels of neural activity, while the ovals are intended to group cell-assemblies encoding for the same class of the percept. Inhibitory network manage a soft winner-wakes-all dynamics (soft-WTA) (i.e. multiple groups are permitted to activate, see Sec.~\ref{subsec:methods:brain_states} for more details). Due to the two areas cooperation, cortical groups encoding for the same class are easily co-activated.
    \textbf{D}) NREM stage mimicking the \textit{Apical Isolation} mechanism: no perceptual input is fed into the network and inter-area cortico-cortical connections are cut. Synaptic STDP is active in intra-area cortico-cortical connections. Each area is thus isolated. Inhibitory network manage an hard-WTA.
    \textbf{E}) REM stage mimicking the \textit{Apical Drive} mechanism: no perceptual input is fed into the network. Synaptic STDP is active in both intra-area and inter-area cortico-cortical connections. The internal input directly determines the output of the network. Inhibitory network manage a soft-WTA. Due to the two areas cooperation, cortical groups encoding for the same class are easily co-activated.
    }
    \label{fig:net_struct}
\end{figure}

This work proposes a two-area thalamo-cortical spiking network that takes inspiration from experimentally grounded hypothesis about the mechanisms underlying brain states and cognition. The structure of the network is based on the hierarchical organisation principle of the cerebral cortex and on the supporting cellular mechanisms proposed by~\cite{aruLarkum2020, larkum2013}, rooted on the combination of feed-forward (coming from lower-hierarchy regions, e.g. sensorial stimuli) and feed-back (coming from other, possibly higher-hierarchy regions, e.g. internal predictions) signals (Fig.~\ref{fig:exp_ground}B, bottom). In turn, this is related to the principles of \textit{apical drive} and \textit{apical isolation}~\cite{aru2020}, and to the role played by neuro-modulators in the expression of different brain states (awake, NREM, and REM) (Fig.~\ref{fig:exp_ground}C, right). Finally, multi-area interaction and activity in different states (and specifically isolation between areas in NREM stage) have been implemented based on experimental observation in~\cite{sarasso2014} (Fig.~\ref{fig:exp_ground}A). This network is capable of expressing both REM- and NREM-like dynamics integrating the perception of two hemisphere. 
Each modelled area includes thalamic relay (\textit{tc}) and reticular (\textit{re}) neurons in the thalamic layer, as well as pyramidal neurons (\textit{cx}) and two populations of inhibitory interneurons in the cortex. The latter include fast spiking neurons (\textit{f-inh}), characterised by strong excitatory to inhibitory connections, and slow spiking neurons (\textit{s-inh}), characterised by weak excitatory to inhibitory connections. Inclusion of both types of inhibitory neurons in our model is crucial to simulate and regulate REM sleep-like dynamics.  All neurons are conductance-based Adaptive Exponential Integrate and Fire (AdEx), modelled through the NEST simulation engine~\cite{nest}, more details in Methods Sec.~\ref{subsec:methods:network}. The two areas receive a sensorial signal of two complementary and partially overlapping inputs, projected towards the thalamus through independent Poissonian spike trains. This specific input mechanism is inspired by, and to some extent emulates, a retinic signal (see Methods Sec.~\ref{subsec:methods:preprocessing} for more details on input preprocessing of the foveal-like protocol). However, all memory storage and recall mechanisms are quite generic, because each training image is simply associated with patterns of higher activity in \textit{th} populations and therefore generic and extendable to other sensorial modalities or any cortical area reached by projection of a specific thalamic nucleus. In addition, the extension to a multiplicity of areas working in parallel (that is, at the same level of hierarchy) seems quite immediate.

\subsection{Simulated cognitive states and comparison with experimental observations}

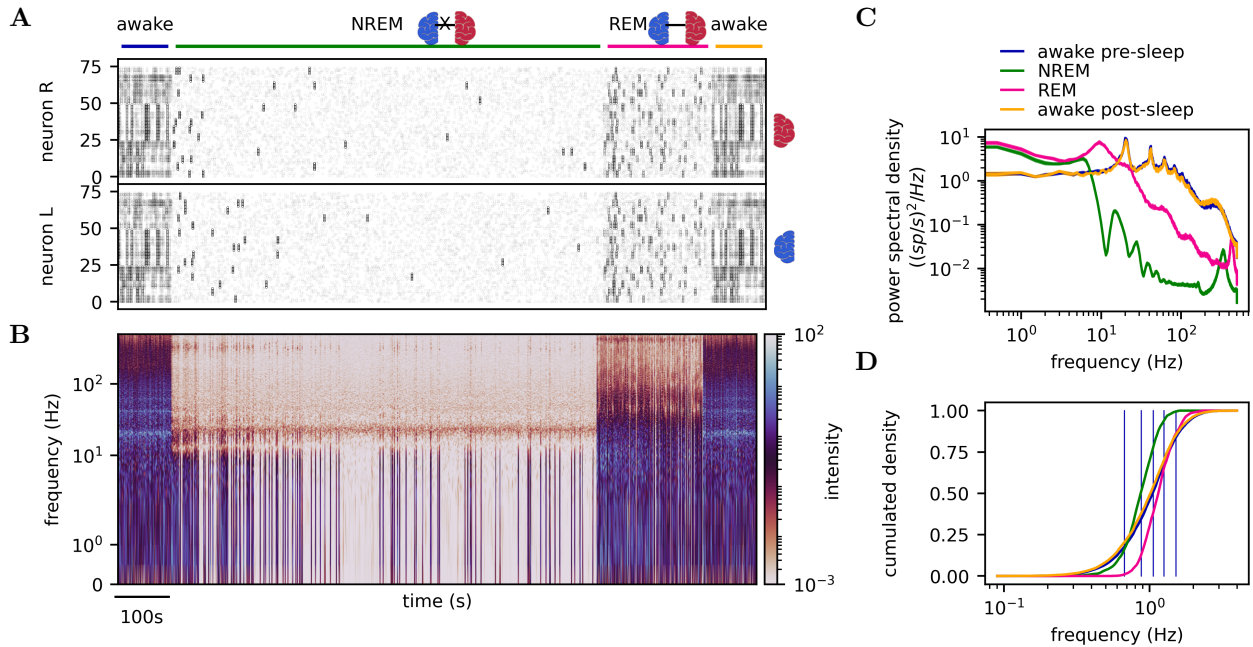
\begin{figure}[!t]
    \centering
    \includegraphics[width=\textwidth]{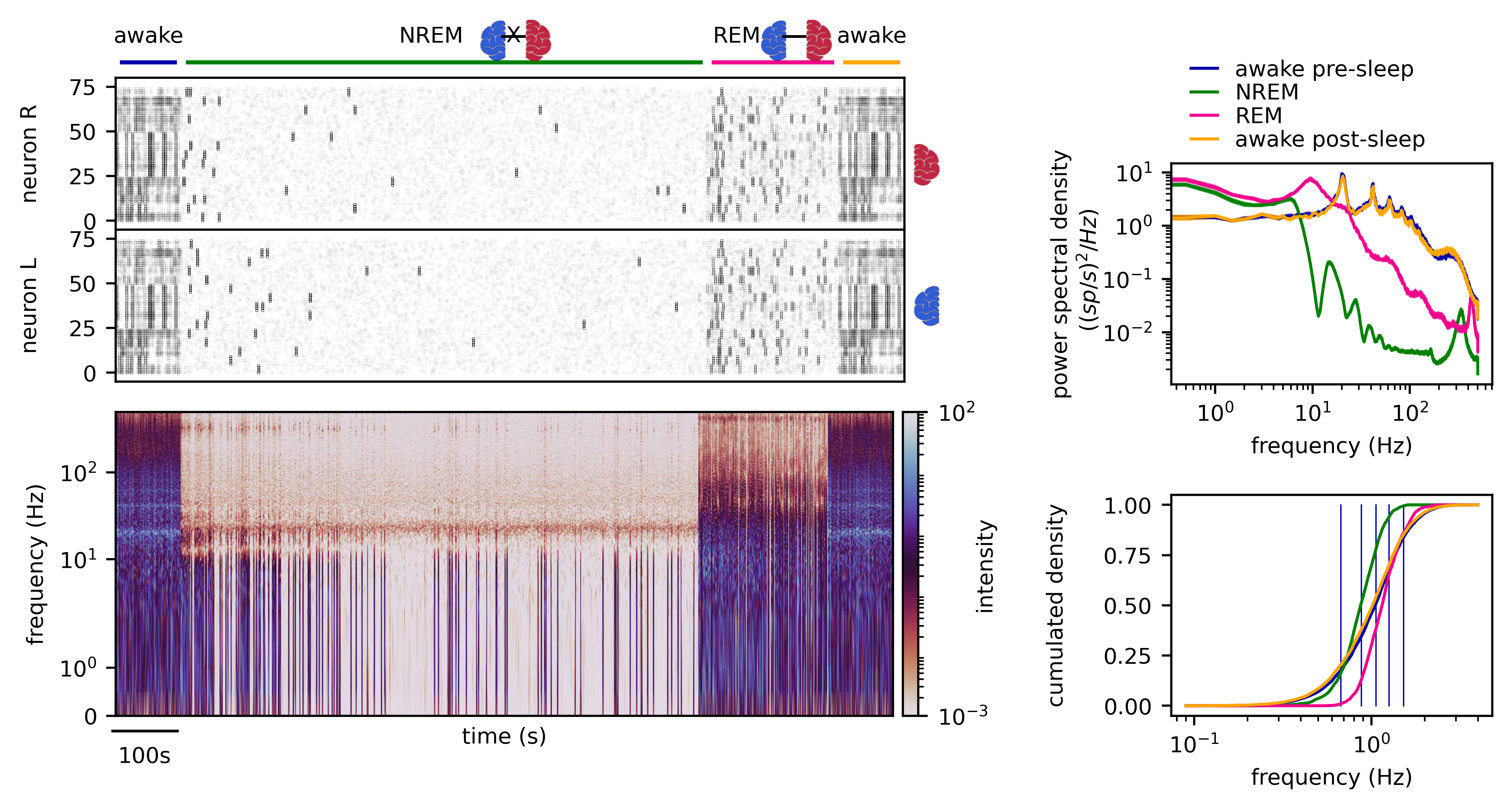}
    \put(-475,240){\textbf{A}}
    \put(-475,120){\textbf{B}}
    \put(-155,240){\textbf{C}}
    \put(-155,110){\textbf{D}}
    \caption{
    \textbf{Biological plausibility of the simulated network: comparison with experimental observations.}
    \textbf{A)} Rastergram showing the activity of the excitatory cortical network throughout waking state - NREM-sleep - REM-sleep.
    \textbf{B)} Time-resolved power spectrum of the simulated network activity during the states experienced by the network; to be compared with fig. 1A by Watson et al.~\cite{watson2016}.
    \textbf{C)} Power Spectral Density across simulated brain states, to be compared with fig. 3B in~\cite{gonzalez2019}.
    \textbf{D)} Cumulative firing rate distribution of excitatory cortical neurons, to compare with Watson et al fig. 2A and Golosio et al. fig. 2A~\cite{watson2016, golosio2021}. A slight reduction in the post-sleep awake distribution can be observed comparing the blue and yellow curve.
    }
    \label{fig:experimental_comparison}
\end{figure}

First, we question the biological plausibility of the different ``cognitive states'' expressed by the proposed network by analyzing the network activity in both time and frequency domains. Fig.~\ref{fig:experimental_comparison}A depicts the rastergram of the network whereas the time-resolved fast Fourier transform-based power spectrogram and the power spectral density (PSD) of such activity are provided in Fig.~\ref{fig:experimental_comparison}B-C.  Finally, the cumulative distribution of the firing rate of excitatory cortical neurons in the three simulated cognitive stages is shown in Fig.~\ref{fig:experimental_comparison}D. These measures are comparable with the ones experimentally observed in~\cite{gonzalez2019} and~\cite{watson2016} supporting the biological plausibility of the implemented thalamo-cortical network. More details about how these stages are implemented and emerge from the network can be found in Methods (Secs.~\ref{subsec:methods:brain_states} and~\ref{subsec:methods:apicalPrinciples}).

Specifically, during the NREM stage, the network expresses low-frequency synchronous activity, typical of slow oscillations (Fig~\ref{fig:experimental_comparison}A) with most of the power concentrated in the $\delta$ band (i.e. $0-4$Hz) with a sharp cut-off at about $10$Hz and a peak at $0.5$Hz (Fig.~\ref{fig:experimental_comparison}C). Furthermore, during this stage, the activity of the network decreases in frequency and amplitude over time as a result of the homeostatic effect of this sleep stage (Fig.~\ref{fig:experimental_comparison}B), coherently with experimental measures~\cite{Tononi2014}. Firing rates associated with these stages are lower and less sparse than the ones observed in others.\\
The REM stage, on the other hand, is characterized by a high-frequency and asynchronous behavior, its activity-power is distributed mainly between $0-20$Hz and presents a peak around $9.5$Hz, in the proximity of the $\theta$ band (i.e. $4-8$Hz), then followed by a power-law decay (Fig.~\ref{fig:experimental_comparison}C). The neuronal firing rate distribution is sharper than the one observed in the awake stage with a comparable median value.\\
Finally, the awake state exhibits a sharp peak at $20.5$Hz in the $\beta$ band (that is, $13-30$Hz), which is in agreement with what was expected for the task-engaged brain power spectrum~\cite{PATHANIA202118}.  It is possible to observe a wide variety of firing rate activity expressed by the network and a small shift lowering the network firing rate in the awake stage after the sleeping phase.

\subsection{Homeostatic and associative effects of NREM and REM activity on the internal structure of the network}

Once demonstrated the biological plausibility of the simulated dynamics of the three cognitive states in terms of spectral features, we measure the effects of two NREM-REM sleep cycles on the internal structure and activity of the implemented thalamo-cortical network.

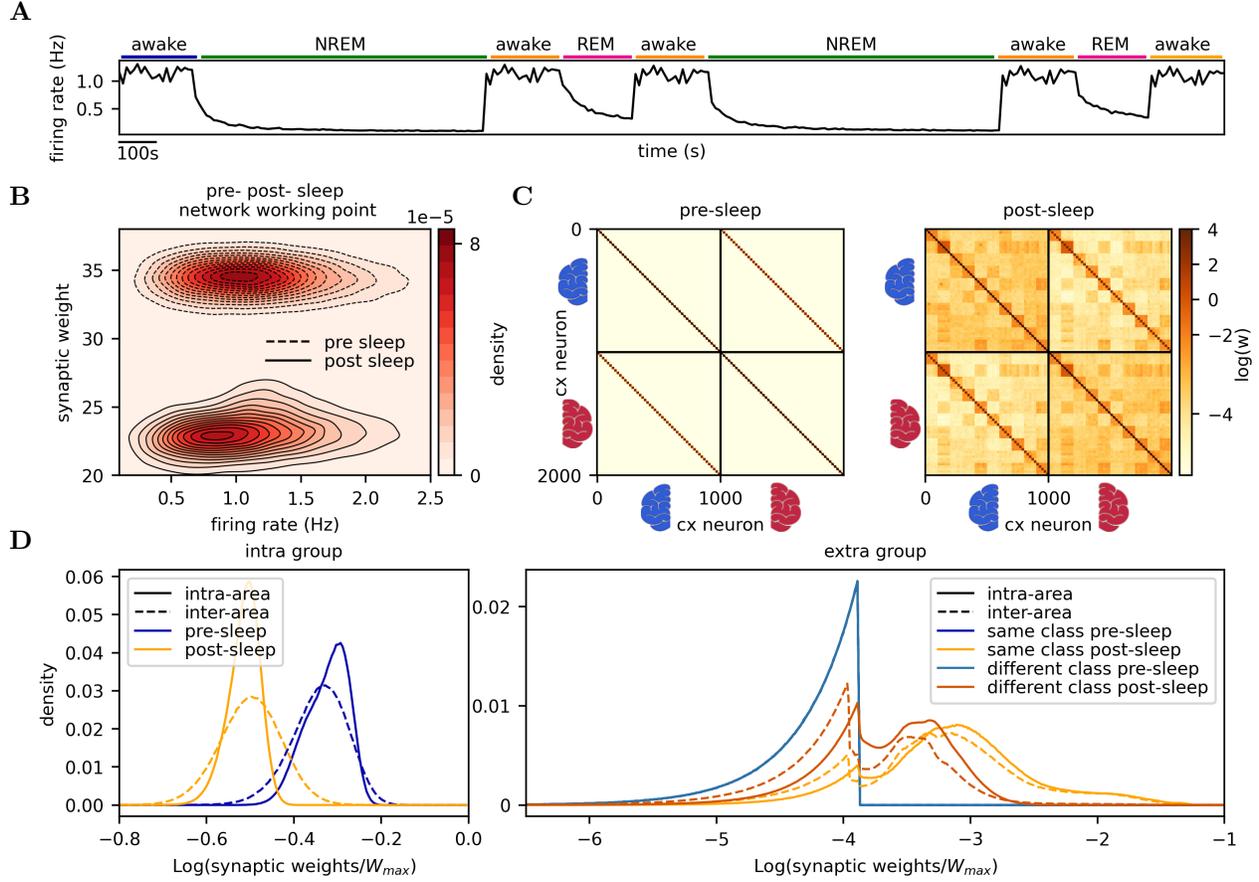
\begin{figure}[!t]
    \centering
    \includegraphics[width=\textwidth]{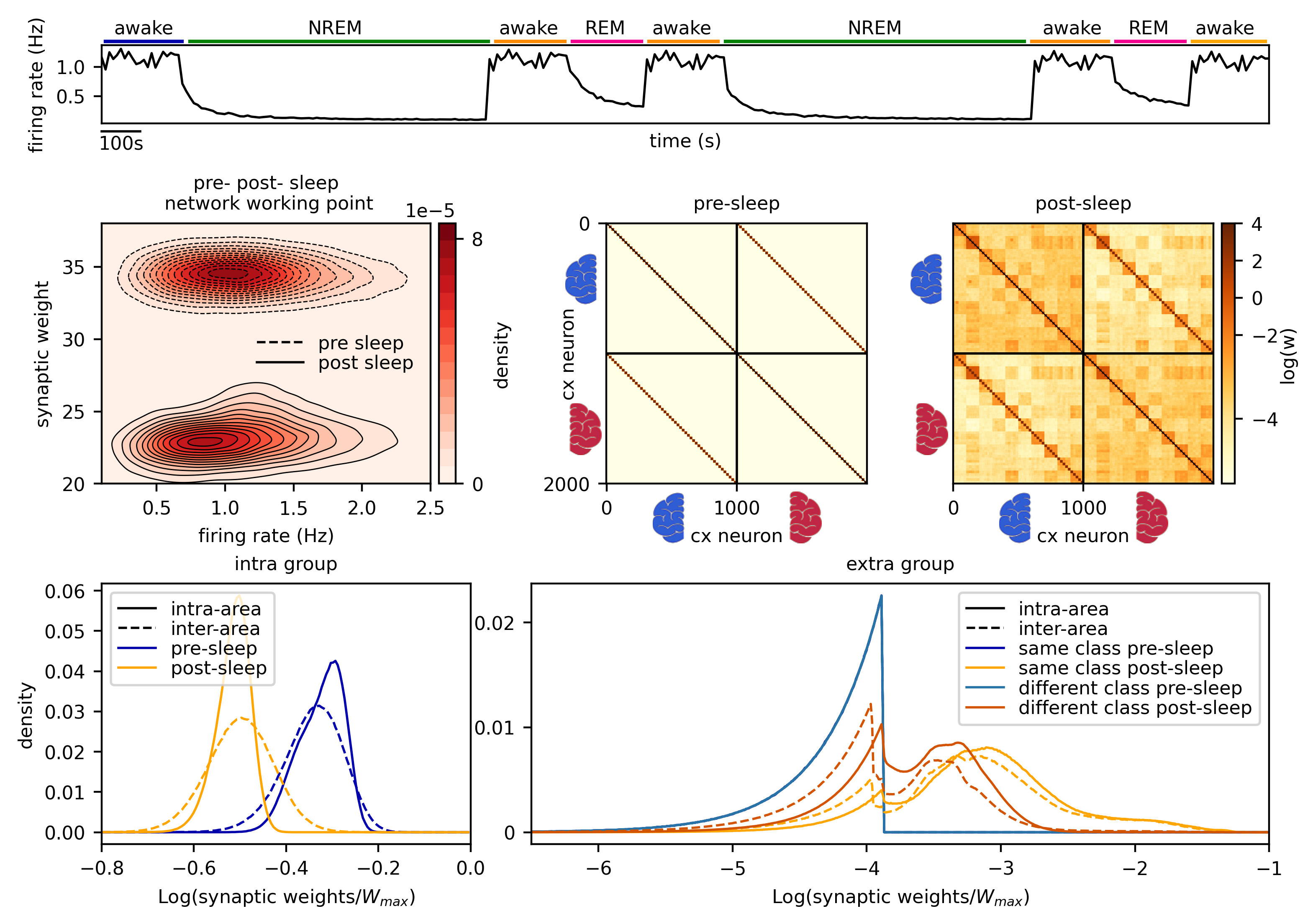}
    \put(-475,330){\textbf{A}}
    \put(-475,260){\textbf{B}}
    \put(-475,130){\textbf{D}}
    \put(-285,260){\textbf{C}}
    \caption{
    \textbf{Effects of NREM and REM activity on the internal structure of the network}
    \textbf{A)} Mean firing rate of excitatory cortical neurons within two awake-NREM-REM sleeping cycles. To be compared with~\cite{watson2016} fig.3B and~\cite{golosio2021} fig. 2D. 
    \textbf{B)} Homeostatic and sharpening effects of sleep-like activity on mean recurrent cortico-cortical excitatory synaptic weights and mean cortical excitatory neurons firing rates. Data collected over $10$ independent training subsets.
    \textbf{C)} Sleep effect on the cortico-cortical synaptic weight matrix. The matrix shows the network trained over $5$ examples per class ($10$ classes), $20$ neurons per example per area before the sleeping phase (left) and after two NREM-REM cycles (right). It is organized so that neurons stimulated by the same example are close together and groups associated to the same class lie on the diagonal. Moreover, the two areas are depicted with the same order (e.g. neurons trained over the first example are numbered $0-20$ for the right area and $1000+0 - 1000+20$ for the left one. Before sleep, the main diagonal depicts intra-group (i.e. same example) intra-area strong connection whereas the secondary diagonals depict the inter-group inter-area strong connection. This matrix is shaped during the training phase, storing orthogonal memories. After the sleeping phase, on the other hand, two effects can be observed: a general reduction of strong connections (homeostasis) and the strenghtening of synapses connecting groups trained over different examples belonging to the same class (association). 
    \textbf{D)} Sleep effect on the cortico-cortical synaptic inter-area (dashed) and intra-area (solid) distributions. \textit{Left}: synapses connecting neurons trained over the same example. The sharpening of the distribution and the reduction of the average weight due to the sleeping phase is shown in both inter-area and intra-area distributions. \textit{Right}: distribution of synapses connecting neurons trained over different examples belonging to the same class and to different class. Before the sleeping phase, both distributions overlap in an exponentially decaying distribution: all memories are stored orthogonally, no extra-group connection is shaped. After the sleeping phase, on the other hand, neurons coding for similar memories are grouped toghether for both intra-area and inter-area connections: this is shown by the separation of the synaptic distributions connecting neurons trained over the same class (yellow) and different class (brown).
    }
    \label{fig:internal_effects}
\end{figure}

Fig.~\ref{fig:internal_effects}A shows the mean firing rate of excitatory cortical neurons within these cycles, to be compared with the experimental measures~\cite{watson2016}. A direct comparison of the joint distributions for firing rates and synaptic weights before (dashed lines) and after (solid lines) sleep is then reported in Fig~\ref{fig:internal_effects}B. The homeostatic effect of sleep can be directly appreciated through the strong reduction of recurrent cortico-cortical synaptic weights, and also through a decrease in the mean firing rate of neurons during classification in the awake state.
The same plot also hints for the reduction in the power consumption within the cortical population as a combined effect of the decrease in the firing rates and synaptic strenght (see Sec.~\ref{sec:method:power} in Methods for the details on power estimate).

Concerning the effects of sleep on the internal structure of the network, recurrent cortico-cortical synaptic distributions are shown and compared before and after the sleep phase in Fig.~\ref{fig:internal_effects}C,D. Specifically,  Fig.~\ref{fig:internal_effects}C shows the recurrent synaptic matrix after training on $5$ examples per class. This is organized so that neurons trained over the same example (group) are close to each other and groups trained over examples belonging to the same class are sequential. Both areas are organized in the same order. This pre-sleep matrix is shaped during the training phase, storing orthogonal memories: no synaptic connection between neurons coding for different examples is shaped. After the sleeping phase, on the other hand, two effects can be observed: a general reduction of strong connections (homeostasis) and the strengthening of synapses connecting groups trained over different examples belonging to the same class (association), as depicted by bigger squares on both the main diagonal (intra-area connections) and the secondary ones (inter-area connections).  In Fig.~\ref{fig:internal_effects}D both these effects are quantitively measured. The homeostatic effect of sleep is demonstrated by a reduction of the weight distributions normalized by $W_{max}$ of synapses connecting neurons coding for the same example (left), for both intra-area and inter-area connections. In fact, the median of this distribution is reduced from $0.48$ to $0.31$ for the intra-area connections and from $0.46$ to $0.31$ for the inter-area connections. Furthermore, a sharpening effect can be observed in intra-area synaptic connections: in fact, the standard deviation is also reduced from $0.05$ to $0.03$. The effect on inter-area distributions, on the other hand, is not significant, remaining stable at $0.06$. The associative effect of sleep manifests in synapses connecting neurons coding for different examples, as depicted in Fig.~\ref{fig:internal_effects}D, right.  After the sleeping phase, on the other hand, the distributions of synapses connecting neurons coding for examples belonging to the same class (more similar, yellow lines) and those connecting neurons coding for different classes (brown lines) are both increased and still separated. Indeed, the former is stronger, depicting the formation of macro-groups of neurons coding for different examples belonging to the same class, demonstrating an associative effect of deep-sleep dynamics towards abstraction and generalization of learnt memories. This effect can be consistent in both intra-area and inter-area connections. To quantitatively measure this separation in intra-class and inter-class synaptic weight distributions, we computed the p-value associated with a two-sample Kolmogorov-Smirnov test \cite{hodges1958significance} in such distributions: it is $<10^{-16}$ for intra-area distributions and $<10^{-16}$ for inter-area distributions. In both conditions, the null hypotesis is rejected proving that intra-class and inter-class synaptic connections evolve differently during the sleeping phase.

\subsection{Beneficial energetic and cognitive effects of sleep}
\label{subsec:results:beneficial_effects}

Finally, we discuss how the effects of sleep on the internal structure of the network presented in the previous section affect the awake state for what concerns the energetic consumption and classification performance. We evaluated the effects of 2 cycles of NREM-REM sleep. The network has been trained on $5$ examples per each of $10$ handwritten digits classess of the MNIST dataset \cite{mnist}. The effect of sleep on the classification performance has been evaluated on $210$ images of testing dataset balances on the ten digit classes. Also, we estimated the reduction in power consumption of the network when performing the post-sleep classification task.  Sec.~\ref{sec:method:power}.

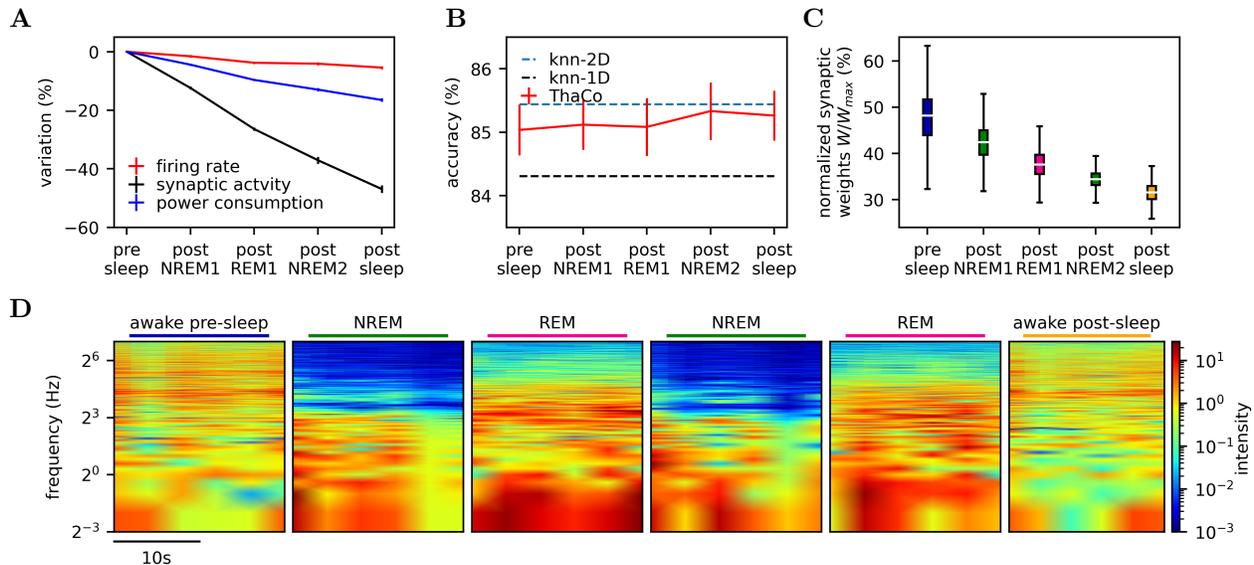
\begin{figure}[!t]
    \centering
    \includegraphics[width=\textwidth]{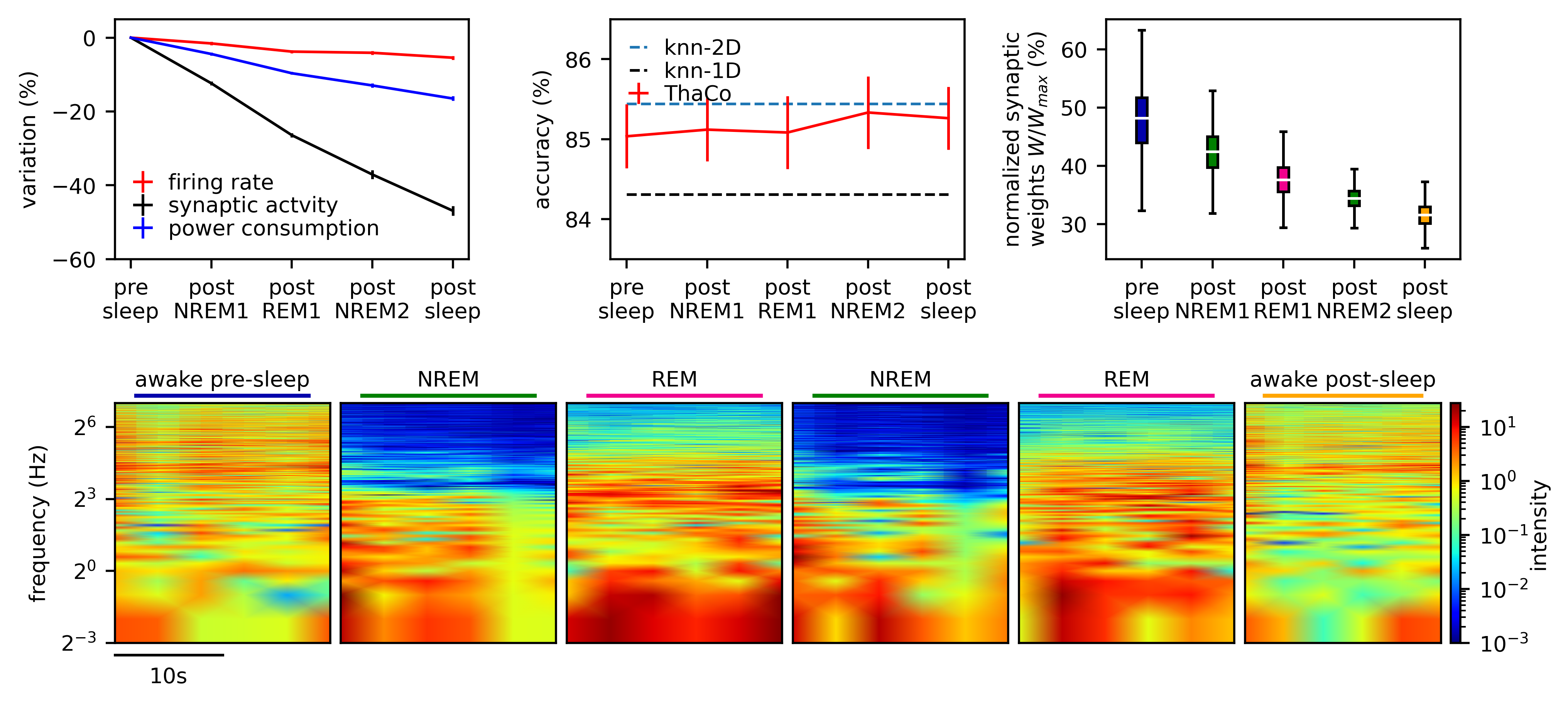}
    \put(-475,210){\textbf{A}}
    \put(-310,210){\textbf{B}}
    \put(-475,100){\textbf{D}}
    \put(-175,210){\textbf{C}}
    \caption{
    \textbf{Beneficial cognitive and energetic effects of two-cycle sleep-like dynamics.}
    \textbf{A)} Network energetic consumption when performing and awake classification task. It shows the percentage change in network mean power consumption (blue, see Methods Sec.~\ref{sec:method:power}) obtained from the mean firing rate (red) and the synaptic (black) activity variations of the network performing the same classification task after each sleeping state.
    \textbf{B)} Network classification performance across brain states. It shows the classification accuracy of the network trained on $5$ examples per category ($10$) on the MNIST dataset after each sleeping stage. The dashed lines (green, blue) show the comparison with the KNN4 classifier (1D: mono-areal, 2D:bi-areal, see Methods Sec.~\ref{subsec:methods:KNN}). Classification accuracy is stable, close to the theoretical upper limit. Averages are performed over $40$ balanced independent training subsets.
    \textbf{C)} Homeostatic effect of each sleeping phase on cortico-cortical intra-group excitatory synaptic weights. Both downscaling and sharpening of the synaptic distribution are shown, leading to network activity and energetic consumption reduction.
    \textbf{D)} Time-resolved power spectrum of the simulated network activity during the first $20s$ of each awake and sleep stages.
    }
    \label{fig:beneficial_effects}
\end{figure}

First, as shown in Fig.~\ref{fig:beneficial_effects}A, the total power consumption of our small model reduces of $17\%$ (see Methods Sec.~\ref{sec:method:power}). This is the result of the decrease in synaptic activity and the mean firing rate that reduce of $50\%$ and $5\%$, respectively. This is achieved by the decreasing and sharpening of the cortico-cortical synaptic weight distribution during sleep and thus post-sleep neural activity. Additionally, the low relative reduction in mean firing rate compared to synaptic change is consistent with expectations based on SHY hypothesis~\cite{CIRELLI2017}. Notably, in this small network all memories are involved and this results in a strong energetic effect. On larger networks, already exposed to large training data sets during previous awake-sleep cycles, the homeostatic effect would be limited to memories learned during the last awake cycle or strengthened by reactivation. However, small power consumption reductions induced by each sleep cycle would exponential compound over days and years of awake/sleep cycles. The gain in energetic consumption does not affect the network classification performance, as shown in Fig.~\ref{fig:beneficial_effects}B. The network classification performance is compared with the K-Nearest-Neighbour algorithm ($K=4$) neighbors computed both independently on the two portions of the image (KNN-1D) or grouping together both features (KNN-2D), which provides an upper limit to the classification accuracy of this type of network (see~\cite{golosio2021} and Methods Sec.~\ref{subsec:methods:KNN}). Performances are always higher than performances of a KNN-1D. After the sleeping phase, they stay near the upper limit defined by KNN-2D in which the two hemispheres are cooperating to classify the provided input.
Comparing the effects of multiple NREM-REM cycles, instead of a single one, Fig.~\ref{fig:beneficial_effects} demonstrates that the network reaches a better operating point, performing the cognitive tasks with performances near to optimality while reduces the required energetic cost. Specifically, Fig.~\ref{fig:beneficial_effects}D shows the spectrogram of the first $20$s activity of the network for each cognitive stage in two-cycle sleep-like dynamics, whereas Fig.~\ref{fig:beneficial_effects}C depicts the homeostatic effect after each stage on the recurrent cortico-cortical synaptic connections.

\section{Discussion}
\label{sec:discussion}

The beneficial effects of sleep on cognitive performance are experimentally evident~\cite{Killgore2010}, even though the mechanisms involved are yet to be understood in their details. Mechanisms that reduce the synaptic activity and, consequently, firing rates provide physiologic advantages when performing a cognitive task that can be measured using energy to solution metrics. Association mechanism and reorganization of memories can increase the cognitive ability toward generalization and abstraction but are associated to higher energetic costs and must be therefore kept under control. In this work, we limited to study these aspects, while we neglected other important topics like the sleep-induced restoration of ionic equilibrium and nutrients and other reparative processes.
Specifically, we explore the network reorganization through a computational model that captures several aspects of biological networks and brain-state induction, in which cognitive functions and sleep-induced rhythms interact.
We consider a spiking network trained to store and classify handwritten digits in a target-based fashion~\cite{capone2019, muratore2021target, capone2021error, capone2022burst}. This strategy allows us to carefully choose the initial internal representation of learned memories, and to evaluate in a controlled way their evolution during sleep phases. This is important to understand how sleep-like dynamics propagates information from one region to another~\cite{capone2017speed}, and how this can be used to efficiently reorganise stored knowledge. In particular, we have extended the thalamo-cortical spiking model, recently published~\cite{capone2019, golosio2021}, already able to express awake and NREM states, that demonstrated sleep induced cognitive gains during classification of noisy images, to a bi-area model capable of experiencing REM-like sleep and multiple brain state cycles. At the same time, we aim to characterize such brain states activity by comparing observables such as power spectral density and neuronal mean firing rate with experimental findings consolidated in the literature~\cite{gonzalez2019, watson2016} and to show the effect of sleep on network energetic and classification performance during a digit classification task. The network architecture has been designed to capture the binocular visual information integration, which takes place in the primary visual cortex during awake-learning through the thalamo-cortical feed-forward stream, making use of the apical amplification principle to encode such information into a bi-area cell-assemblies network. Furthermore, the principles of apical isolation/drive are used to set the cortical neuromodulation to switch between awake and NREM / REM states and thus to reproduce cortical activity generally in agreement with the observations of~\cite{massimini2014} on the spread of cortico-cortical activity during awake and NREM/REM sleep. As a result the network has been found able to perform learning and sleeping cycles holding classification performance close to the accuracy upper-limit for such a network (see Methods Sec.~\ref{subsec:methods:KNN}), decreasing at the same time mean firing rate and synaptic power consumption and keeping cortical activity coherent with biological recordings (Figs.~\ref{fig:experimental_comparison} and~\ref{fig:beneficial_effects}). Furthermore, we have shown the ability of sleep to shape cortico-cortical connectivity resulting in the down-scaling and sharpening of the synaptic distribution, and thus of the mean firing rate, as well as the creation of novel multi-area associations. These results allow us to argue that NREM and REM sleep cycles are useful in such a network to reduce energy consumption while reorganizing the synaptic structure of the network. 
As a future perspective, we propose to better characterise the opposite effects of homeostasis and association in terms of the entropy of the synaptic matrix: whether entropy reduction occurs at the synaptic level as a consequence of synaptic downscaling and sharpening, sleep-induced associations are collaborating with the learning process to produce entropy, thus extending the state space explorable by the network. It will be interesting to investigate under which conditions the sum of homeostatic and associative contributions to synaptic entropy is positive or negative.
Last, it is important to notice that the structure of ThaCo, in synergy with the apical amplification, allows the network to associate particular percepts with particular thalamo-cortical patterns, endowing the network with a set of independent neural groups. The groups can interact during both waking and sleeping phase, autonomously evoking specific trajectories in the memory space (for a visual illustration, see the figures in SM Sec.~\ref{subsec:SM:WTA-mechanisms}). So, we can say that ThaCo provides a framework in which mechanisms (in this case the neural correlates of memories) exist intrinsically~\cite{tononi2015} and can influence other mechanisms during both awake, NREM (see~\cite{capone2019, golosio2021}) and REM phases through exogenous or endogenous stimuli, thus giving rise to cause-effect complex that consolidate in new accessible mechanisms. In addition, a fundamental ingredient of ThaCo is integration of information among hemispheres: as in awake and REM sleep Apical Amplification/Drive support cooperation between brain areas, during NREM sleep Apical Isolation acts in such a way that information is segregated in a single brain area, hence reducing the state space experienceable by the network. For these reasons, we argue that ThaCo could provide a framework for testing Integrated Information Theory (IIT)~\cite{tononi2015}, directly or through measures of the Perturbational Complexity Index (PCI)~\cite{massimini2014}. 

Further optimizations and integration of this work can be done in the perspective of large scale simulations. Based on previous studies of Slow Waves Activity in both spiking~\cite{pastorelli2019} and mean-field~\cite{capone2022infer} large scale cortical simulations, we could improve the present model to exploit the interplay between Slow Waves propagation and cognitive tasks in larger networks.

\section{Methods}
\label{sec:methods}

\subsection{Network}
\label{subsec:methods:network}

The model is based on a 2-area 2-layer spiking network, as illustrated in Fig.~\ref{fig:net_struct}. The two areas receive in input an independent perceptual signal; the thalamic layers are not reciprocally connected, whereas the cortical layers are connected. The first layer in each area, the thalamus, is composed of a population of excitatory neurons and a population of inhibitory neurons; the cortical layer is composed of a population of excitatory neurons connected to 2 independent inhibitory populations: one, fast-spiking, with a strong excitatory to inhibitory synaptic connection, and a second, slow-spiking, with a weak excitatory to inhibitory synaptic connection.  The described network has been implemented in NEST 3.1~\cite{nest, Aeif_cond_alpha} and is made of conductance-based adaptive exponential (AdEx) neurons (see SM Sec.~\ref{subsec:SM:adexNeuron} for more details on neuronal and synaptic dynamics).  All excitatory-to-excitatory synapses are plastic within the training phase, none are plastic in the classification phase, whereas only cortico-cortical synapses are plastic during sleep-like phases. All synapses connecting excitatory to inhibitory neurons and vice versa are not-plastic. Plasticity is described by an STDP dynamics (see SM Sec.~\ref{subsec:SM:STDP}).

\subsection{Fast and late inhibition}
\label{subsec:methods:fastLateInhibition}

Within the network, two independent inhibitory populations are identified, acting on different time scales: the fast-spike population is able of fast responding, regularly firing up to $500$Hz for tens of milliseconds, while the slow-spike population responds at around $10$Hz. This choice is motivated because it allows the network to provide time-increasing inhibition to cortical neurons. Therefore, fast-spike neurons are intended to set the excitatory activity level and achieve \textit{soft winner-take-all dynamics} (\textit{soft-WTA}) while slow-spike neurons are used mainly to make the network able to exit memory attractors during REM sleep (see Methods Sec.~\ref{subsec:methods:brain_states} for the different mechanism to exit memory attractors during NREM sleep). It is worth noting that fast and slow behaviors are emulated acting on the inhibitory population excitability only, in order to avoid neural parameters fine tuning. Specifically, a slow and fast response is achieved by properly tuning up the weights of excitatory synapses projecting toward the two inhibitory populations.

\subsection{Combination of contextual and perceptual signals during the training phase} 
\label{subsec:methods:combinationContextPeception} 
During the training phase, images of handwirtten digits from the MNIST dataset \cite{mnist} are provided to the network for $1.5$s together with a contextual example-specific signal stimulating a subset of 20 cortical excitatory neurons. During the classification phase, new, unseen inputs are presented to the network for $1$s without any contextual information (see SM sec.~\ref{subsec:SM:Parameters} for more details).
During the training phase, a set of memories are encoded in groups of cortical neurons (\textit{cx}) according to an unsupervised protocol through STDP mechanisms (see SM Sec.~\ref{subsec:SM:STDP}); during the presentation of the training example, \textit{cx} neuron receive both the bottom-up perceptual signal from the thalamic layer of the network and a time-specific contextual signal which provides to the cortical layer a pure temporal label, with no info about the label of the class the novel example belong to. This emulates the existence of a time-specific status in other areas in the brain collected by the apical tuft of the \textit{cx} neuron. Thanks to the action of the STDP mechanisms, those neural groups that are simultaneously activated in multiple areas by the time-specific contextual label develop two sets of synapses: a) a set of strong intra- and inter-areal synapses among themselves, and b) strong connections among the thalamic active pattern and the group of \textit{cx} neurons facilated by the time-specific context.  It should be noticed that performance in classification accuracy is measured on the bases of the \textit{cx} neurons that are trained using the unsupervised protocol. During the classification phase, new previously unseen perceptual inputs are provided to the network in the absence of any contextual signal: these inputs activate neurons associated with similar learnt memories, stored at the cortical level.

\subsection{Moving between brain states}
\label{subsec:methods:brain_states}

In order to switch between brain states, the cortical network is provided with both a thalamic (perceptual) input and a cortical Poissonian noise, emulating an non-specific stimulus from external brain areas. In addition to this,  in order to reproduce a neuromodulation effect similar to the one responsible for apical amplification / drive/ isolation, the cortico-cortical excitability is modulated by changing neural Spike-Frequency Adaptation (SFA), cortico-cortical excitation and inhibition level.

\paragraph{Awake}
In the training phase, contextual signal and thalamic feature-specific stimulus are provided to the cortical network, and all cortico-cortical connections are active and plastic (Fig.~\ref{fig:net_struct}A). This setting should be compared with the Apical Amplification situation. However, during the awake state, cortical activity is elicited by the visual thalamic stimulus only, memory evocation is managed by soft-WTA which allows the network activity to visit different memory attractors and to take decisions, while plasticity is set off (Fig.~\ref{fig:net_struct}C). After the training phase, the $W_{\mathrm{intra}}$ and $W_{\mathrm{inter}}$ observables are computed. They respectively represent the averages of the intra-area and inter-area cortico-cortical synapses connecting example-specific neural groups. 

\paragraph{NREM} 
To enter the NREM state, excitatory cortical neurons receive a non-specific stimulus at frequency $1$kHz using a homogeneous Poisson noise generator. Compared to the awake state, cortico-cortical inter-area connections are cut, whilst intra-area synapses are up-modulated by the $\frac{\langle W_{\mathrm{CA}} \rangle}{W_{\mathrm{intra}}}$ ratio (see SM Tab.~\ref{tab:parameters} values). The level of SFA is also increased and the synaptic efficacy of both inhibitory populations is decreased, consistently with the apical isolation principle (Fig.~\ref{fig:net_struct}D).
Therefore, potentiated intra-area connections allow the network to exhibit cell-assemblies up-states dynamics controlled by soft-WTA, sustained by fast-spike inhibitors population, and SFA which leads network out of memory attractors. Slow-spike population here only plays a role during cell-assemblies co-activation, definitively avoiding the chance of network catastrophic forgetting.
After each NREM phase, the network can be re-awakened, e.g. to evaluate its performance. A novel value for $W_{\mathrm{intra}}$
can be measured. In presence of a dominant homeostatic depression, this value is expected to lower after each NREM stage.

\paragraph{REM}
The dreaming state is achieved providing two different non-specific stimuli, at $7$Hz and $30$Hz respectively, to cortical neurons (see SM Tab.~\ref{tab:parameters} for parameter values). In this case, only inter-area cortico-cortical connections are up-modulated by the $\frac{\langle W_{\mathrm{CA}} \rangle}{W_{\mathrm{inter}}}$ ratio, and the SFA level is decreased, in order to emulate the Apical Drive principle. Inhibition changes differently between populations: fast-spike inhibition is depressed, as well as in the NREM phase, while slow-spike inhibition is potentiated (Fig.~\ref{fig:net_struct}E). Here, the high-rate stimulus promotes low gamma activity, while the slow-rate stimulus is intended to emulate the interaction with the thalamus, which is known to impart theta rhythms to the cortex during biological REM sleep~\cite{DURKIN2019}. As a result, high-rate generator put neurons under threshold, while the slow-rate one orchestrates the cell-assemblies activation dynamics. It is worth noting that while fast-spike inhibitory population provides soft-WTA, slow-spike neurons are fundamental to guarantee exit from up-states, because of low-SFA level. Last, the Spike-Timing-Dependent Plasticity time scale has been reduced by a factor $8$, since it greatly enhances REM association performance.

\subsection{Apical drive/isolation/amplification principles} 
\label{subsec:methods:apicalPrinciples}

The network is capable of expressing different brain states: awake-like state, articulated in a training phase (Fig.~\ref{fig:net_struct}A), when memories are stored in the network, and an awake classification phase (Fig.~\ref{fig:net_struct}C), when new stimuli induce activity related to learned memories;  REM-like state (Fig.~\ref{fig:net_struct}E), associated with an activity showing a peak around $7.2$Hz; a NREM-like state (Fig.~\ref{fig:net_struct}D), characterized by a slow oscillation state at $1.2$Hz. The transition between different brain-state dynamics is implemented by means of brain-state specific neuromodulation (Fig.~\ref{fig:exp_ground}C) and changes in excitatory and inhibitory efficacies (for a more detailed description, see Methods Sec.~\ref{subsec:methods:brain_states}). The three implemented states described in Fig.~\ref{fig:net_struct} take inspiration from the study by~\cite{aru2020} summarized in Fig.~\ref{fig:exp_ground}C. The awake state emulates the \textit{Apical amplification} regime, characterized by a selective amplification of information from external stimuli combined with the integration of input from other internal sources. Neuromodulation that regulates the transition between awake and sleep phases, on the other hand, is emulated by modifying the Spike Frequency Adaptation (SFA) and excitation/inhibition mechanisms. During the REM-like phase, no external perception is provided and an Apical Drive mechanism is emulated: internal input directly determines the output with an active Spike-Timing-Dependent Plasticity (STDP) within cortical neurons both intra- and inter-area. During this phase, information learnt in both areas is integrated. During the NREM-like phase, no external perception is provided and inter-area cortico-cortical connections are cut: an \textit{Apical Isolation} mechanism is emulated, intra-area cortico-cortical connections, on the other hand, are active with STDP. 

\subsection{Preprocessing}
\label{subsec:methods:preprocessing}

As already stated, one of the aim of this work is investigating the cooperation between cortical areas connected through inter-area synapses. To do this, we emulated a few aspects of binocular perception: each area has access to the external perception of only half of the original input, with an overlap in the center. 
The image data set is composed of MNIST handwritten digits~\cite{mnist}, in order to implement a two-area perception feature, we improved the preprocessing algorithm already implemented described in~\cite{golosio2021} (SM, pagg. 5-6). In the current work, the salient features of the image are extracted using Histogram of Oriented Gradients (HOG) whereby the image has been sampled at different space-scales in order to emulate foveal perception. In addition, the dataset has been rescaled, embedding each image into a background frame.

\subsection{K-Nearest-Neighbour Algorithm}
\label{subsec:methods:KNN}

The ThaCo model is able to encode perceptual information into  thalamo-cortical and cortico-cortical synaptic matrix in a very general way, using STDP to capture and condensate temporal correlation, among thalamic pattern activation and cortico-cortical group activity, into new synapses. This provides a framework in which memories written into feature-specific thalamo-cortical connections and cortico-cortical recurrent synapses, developed during the learning phase, are represented as orthogonal cortical states~\cite{golosio2021} promoting at the same time a cooperation among groups coding for those pattern that are first neighbours of the learned ones during a classification supported by a soft-WTA mechanism. In such a way, since cortical group activation is very sensitive to the similarity between thalamic pattern related to memories already learnt and the pattern to be categorized, classification tasks as new-image recognition can be conveniently abstracted as selecting the cortical group whose thalamic activation pattern maximises the dot product with the new thalamic pattern. Therefore, it is quite natural to compare the classification performance of such a network with the weighted KNN algorithm. i.e.\ the one that weights the votes nearest neighbours using the inverse of their euclidean distance with the proposed example. The reason for selecting this flavour of KNN is its similarity to the drive that each neural group would exert in a soft-WTA regime. Aiming to take into account binocular perception and biareal processing of ThaCo, we assessed KNN-performance following two different approaches. The first approach KNN-1D considers two sets of perceptual thalamic features related to the left/right (L, R) hemifields. In the KNN-1D case, the weighted KNN algorithm is applied independently to both sets. The winner class is selected according to:

\begin{equation}
    \mathrm{class}_j = \mathrm{argmin}_i(\frac{d_{i,j}^L}{\sum_{i}^4d^L_{i,j}}, \frac{d_{i,j}^R}{\sum_{i}^4d^R_{i,j}}) \quad ; \qquad  i = 1,...,n_{\text{ex},\text{train}} \, , \,\, j = 1, ..., n_{\text{ex},\text{test}}
    \label{eq:knn1d}
\end{equation}

\noindent
where we decided to consider the votes of up to 4 neighbours.

Therefore, KNN-1D selects as winner the class related to the training example which posses the hemifield, is more similiar to the classification instance.

The second approach is KNN-2D: features from left and right hemispheres are grouped to classify the new image. This is the case of monocular perception, so we consider it as the upper-limit to ThaCo classification performance. Also, we argue that when ThaCo classification accuracy is close to this limit, the L/R visual hemifield information is integrated as best as possible by the thalamo-cortical network.

\subsection{Power Consumption}
\label{sec:method:power}

The energy consumption $p$ of each neuron can be considered to be the sum of three contributions \cite{metabolism}:
the basal metabolic cost (firing rate independent); a second term associated with the production of each action potential and the dissipation for its transmission through the axon to reach all the synapses; finally, a term related to the activation of the synapses, the injection of synaptic currents and other synaptic metabolic costs e.g., needed to produce and recycle neurotransmitters at synaptic junctions:

\begin{equation}
    p = b + e \cdot \nu +  c \cdot \omega
    \label{eq:single_neuron_power_consumption}
\end{equation}

\noindent
where $\nu$ is the neuron specific firing rate and $\omega =\sum_j \nu \cdot W_ j$, the neuron specific synaptic activity ($j$ the index enumerating synaptic outputs). The units of measure for the constants $b, e$ and $c$ are $\text{W}$, $\text{J}$ and $\text{J} \cdot \Omega$, respectively.  Their values depend on the neural type and axon arborization geometry. In the proposed model, we measure in each brain state the average firing rate $\nu_{\mathrm{state}}=\langle \nu_i\rangle_{\mathrm{state}}$ and the total synaptic activity $\omega_{\mathrm{state}}=\langle \omega_i\rangle_{\mathrm{state}}$ (see Fig.~\ref{fig:beneficial_effects}A). It is possible to estimate the total power consumption $P$ in a network composed of $N$ neurons, using:

\begin{equation}
    P_{\mathrm{state}} \approx B + E \, \nu_{\mathrm{state}} + C \, \omega_{\mathrm{state}} 
\label{eq:total_power_consumption}
\end{equation}

\noindent
where $B=N\langle b\rangle$, $E=N\langle e\rangle$ and $C=N\langle c\rangle$. 
The relative cost associated to the three terms can be estimated  to account \cite{metabolism} for $\sim 25\%$, $\sim 45\%$, and $\sim 30\%$ of $P$, respectively. 

The post-NREM, post-REM, and post-sleep power consumption can be compared with the presleep power:
\begin{equation}
    P_{\mathrm{post}} \approx  0.25 \, P_{\mathrm{pre}} + 0.45 \, P_{\mathrm{pre}} \frac{\nu_{\mathrm{post}}}{\nu_{\mathrm{pre}}} + 0.30 \, P_{\mathrm{pre}} \frac{\omega_{\mathrm{post}}}{\omega_{\mathrm{pre}}} 
\label{eq:total_post_sleep_power}
\end{equation}

The observed $5\%$ reduction in firing rate and $50\%$ reduction in synaptic activity translates in the model in a $17\%$ reduction of total power (see Fig.~\ref{fig:beneficial_effects}A). 

In this paper we are estimating the energetic effects of sleep  excitatory neurons. 

\section*{Source code availability}

The code to simulate the model and to reproduce the figures can be found at:
\url{https://github.com/APE-group/ThaCo3}. The AdEx neural model is described in SM Sec.~\ref{subsec:SM:adexNeuron}, while the adopted STDP plasticity is reported in SM Sec.~\ref{subsec:SM:STDP}. Neural and synaptic parameters are finally listed in SM Sec.~\ref{subsec:SM:Parameters}.

\section*{Acknowledgements}

This work has been supported by the European Union Horizon 2020 Research and Innovation program under the FET Flagship Human Brain Project (grant agreement SGA3 n. 945539 and grant agreement SGA2 n. 785907) and by the INFN APE Parallel/Distributed Computing laboratory.

\bibliographystyle{unsrt}
\bibliography{Bibliography.bib}

\newpage
\appendix

\setcounter{section}{0}
\setcounter{equation}{0}
\setcounter{figure}{0}
\setcounter{table}{0}
\renewcommand*{\thesubsection}{\Alph{subsection}}
\renewcommand{\theequation}{S\arabic{equation}}
\renewcommand{\thefigure}{S\arabic{figure}}
\renewcommand{\thetable}{S\arabic{table}}

\begin{center}
\rule{\textwidth}{0.05cm}

\section*{Supplementary Material}
\label{sec:supplMat}

\rule{\textwidth}{0.05cm}
\end{center}

\subsection{Group activity co-activation and integration}
\label{subsec:SM:group_activity}

\begin{figure}[!p]
    \centering
    \includegraphics[width=0.9\textwidth]{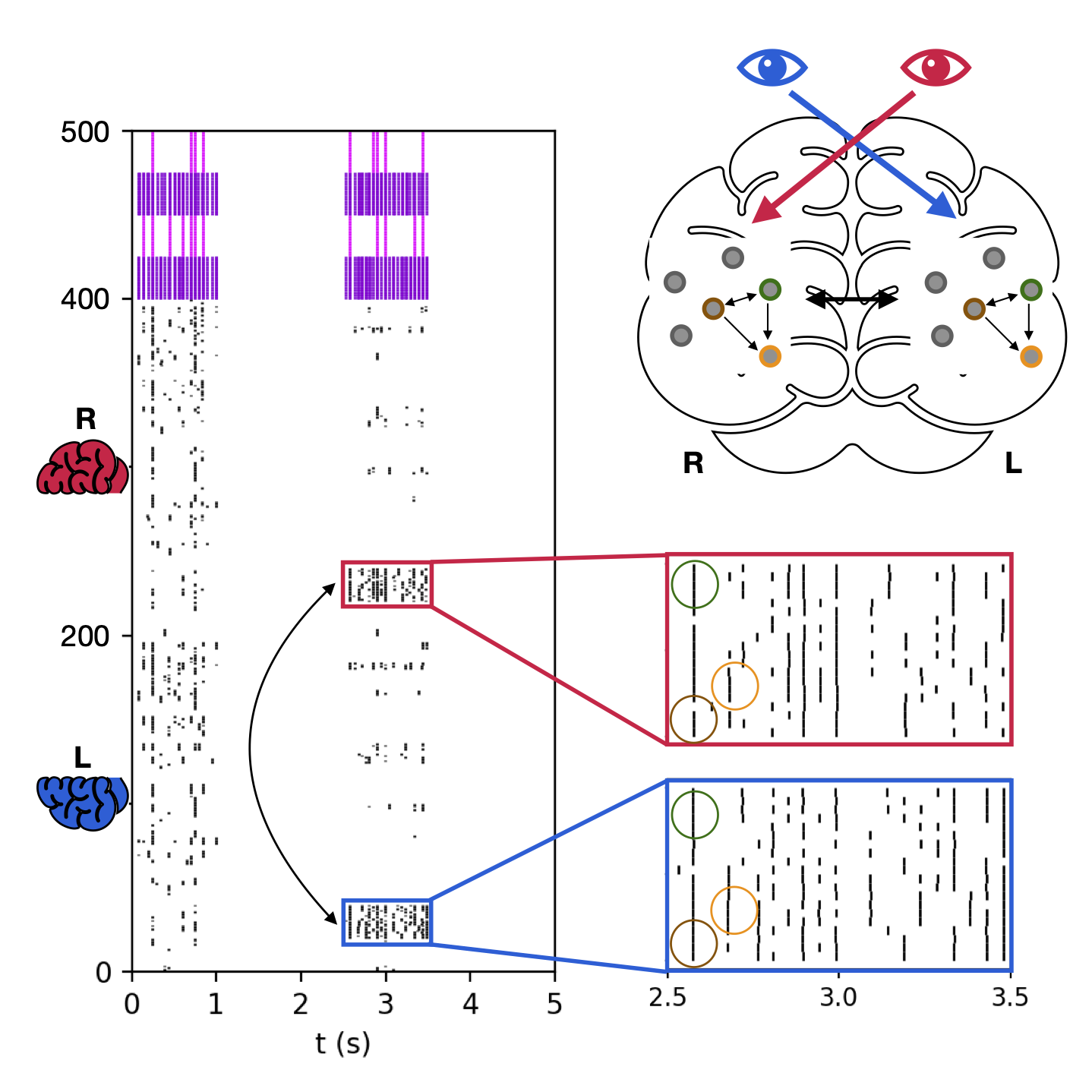}
    \caption{
    \textbf{Neural groups activation: awake}
    Group activity during awake-classification phase. The rastergram on the left shows 5 seconds of the cortical network activity during the awake stage and spatial magnification. Excitatory population is depicted in black, inhibitory population in violet (f-inh) and fuchsia (s-inh). The inset shows activation of neurons belonging to the same class from both areas. It is worth noting that the activity of the two cortical areas is strongly correlated, also in this state groups encoding for the same class collaborate and there are frequent co-activations. On the top-right side, the illustration provides an insight about neural groups collaboration. Grey dots correspond to neural group which refers, through the same colours, to the rastergram just below. Black arrows indicate causal interaction. 
    }
    \label{fig:activity_awake}
\end{figure}

\begin{figure}[!p]
    \centering
    \includegraphics[width=0.9\textwidth]{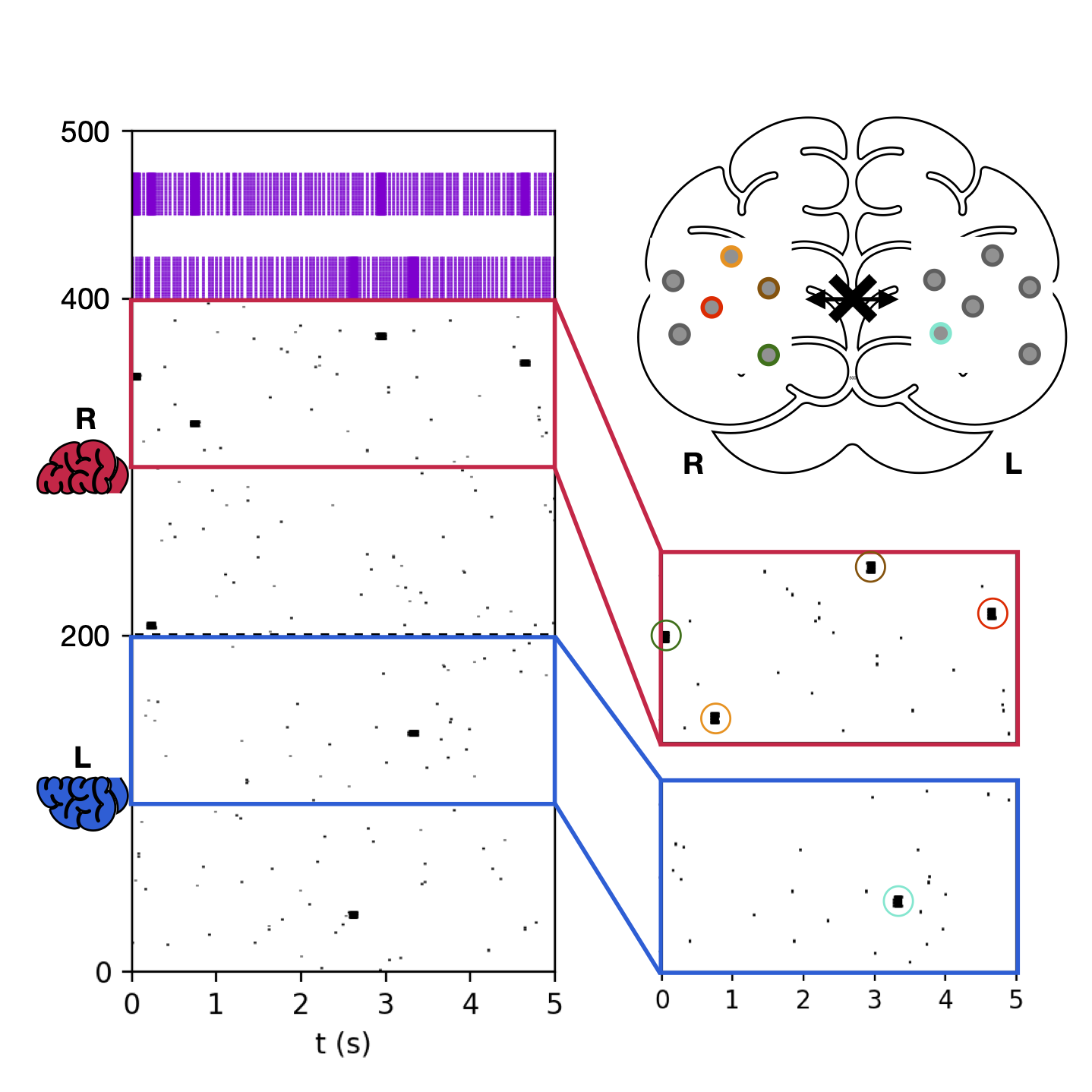}
    \caption{
    \textbf{Neural groups activation: NREM}
    Group activity during NREM phase. The rastergram on the left shows 5 seconds of the cortical network activity during the NREM stage and spatial magnification. Excitatory population is depicted in black, inhibitory population in violet (f-inh) and fuchsia (s-inh). The inset shows activation of neurons belonging to the same class from both areas. It is worth noting that in this state groups encoding for the same class are mostly activated individually and the activity of the two cortical areas is decorrelated. On the top-right side the illustration provides an insight about neural groups activation. Grey dots correspond to neural group which refers, through the same colours, to the rastergram just below.
    }
    \label{fig:activity_nrem}
\end{figure}

\begin{figure}[!p]
    \centering
    \includegraphics[width=0.9\textwidth]{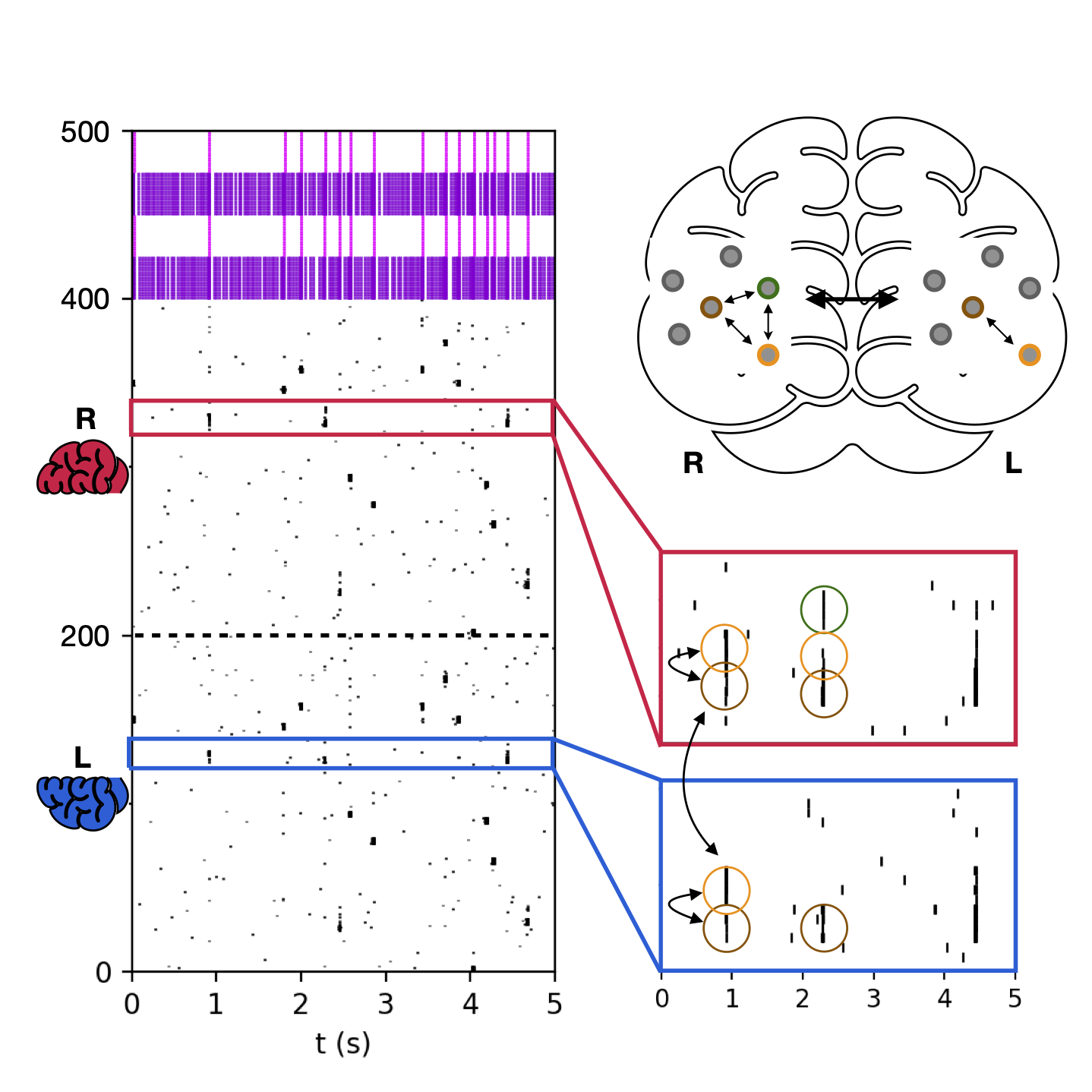}
    \caption{
    \textbf{Neural groups activation: REM}
    Group activity during REM phase. The rastergram on the left shows 5 seconds of the cortical network activity during the REM stage and spatial magnification. Excitatory population is depicted in black, inhibitory population in violet (f-inh) and fuchsia (s-inh). The inset shows activation of neurons belonging to the same class from both areas. It is worth noting that , despite the random exogenous stimuli, the activity of the two cortical areas is strongly correlated, also in this state groups encoding for the same class collaborate and there are frequent co-activations. On the top-right side the illustration provides an insight about neural groups collaboration. Grey dots correspond to neural group which refers, through the same colours, to the rastergram just below.   
    }
    \label{fig:activity_rem}
\end{figure}

Here, we provide further insights into the dynamics of network activity across brain states, with a focus on (co)activation of cortical groups and the inter-area and intra-class integration of information. We show the rastergram during $5$s of cortical activity and a sketch of groups interaction in awake (Fig.~\ref{fig:activity_awake}), NREM (Fig.~\ref{fig:activity_nrem}), and REM (Fig.~\ref{fig:activity_rem}) states.

\subsection{Soft and hard winner take all}
\label{subsec:SM:WTA-mechanisms}

Coherently with what described in~\cite{golosio2021}, the network parameters have been set to induce the creation of WTA mechanisms by emulating the organizing principle of the cortex described in~\cite{larkum2013}. During training, the network is set in a hard-WTA regime (firing rate different from zero only on a selected example-specific subset of neurons), while during classification it works in a soft-WTA regime (i.e., the firing rate can be different from zero in multiple groups of neurons, and the winner group is assumed to be the one firing at the highest rate). Specifically, during the training, we set our parameters so that the thalamic signal alone is not sufficient to cause neurons to spike.

\subsection{AdEx neuron}
\label{subsec:SM:adexNeuron}

The described network has been implemented in NEST~\cite{nest, Aeif_cond_alpha} and is made of conductance-based adaptive exponential (AdEx) neurons defined by the following coupled equations:
\begin{equation}
    \begin{split}
        C_{m}\frac{dV}{dt} &= -g_{L}\left( V-E_{L}\right) + g_{L}\,\Delta_{T}\,e^{\frac{\left(V-V_{th}\right)}{\Delta_{T}}} + I_{\mathrm{system}} - \omega \\
        \tau_{\omega}\frac{d\omega}{dt} &= a\left(V-E_{L}\right) +b\sum_{k}\delta (t-t_{k}) - \omega
    \end{split}
    \label{eq:labelNeuralEquation}
\end{equation}
Here, the first equation describes the time evolution of the membrane potential $V$ and  incorporates a spike-frequency adaptation mechanism through the term $\omega$. On the other hand, the second equation captures the essential features of neuronal fatigue that depend on the number of spikes emitted by the neuron itself in the recent past. Moreover, whenever $V > V_{\mathrm{peak}}$, the membrane potential is set to a reset value $V_{\mathrm{reset}}$.
Here, $\tau_{\omega}$ is the adaptation time constant associated with neuronal fatigue, $C_{m}$ the membrane capacitance, $E_{l}$ the reversal potential, $V_{T}$ the threshold potential, and $\Delta_{T}$ the exponential slope parameter. Moreover, $a$ and $b$ are further adaptation parameters.

The input current from excitatory and inhibitory connections, $I_{\mathrm{system}}$, can be written as:
\begin{equation}
    I_{\mathrm{system}} = g_{\mathrm{ex}}(t)\left(E_{\mathrm{ex}}-V\right) + g_{\mathrm{in}}(t)\left(V-E_{\mathrm{in}}\right)
\end{equation}
where $g_{\mathrm{ex}}$ and $g_{\mathrm{in}}$ are the time-dependent excitatory and inhibitory synaptic conductances, respectively, shaped according to \textit{alpha functions}.
Assuming that a spike occurs at time $t_{s}$, the alpha function for synaptic conductance is defined as
\begin{equation}
    g(t) = \begin{cases} w\frac{(t-t_{s})}{\tau_{s}}e^{-(t-t_{s})/\tau_{s}} & \mbox{if } t>t_{s} \\ 0 & \mbox{if } t<t_{s} \end{cases}
\end{equation}
This conductance has a gradual rise and slow decay, reaching its peak at $t_{\mathrm{max}} = \tau_{s}$. According to this model, the instantaneous current injected by an incoming excitatory synapse spiking at $t_{s}=0$ is then:
\begin{equation}
    I(t) = w \frac{t}{\tau_{s}} e^{-t/\tau_{s}}\,(E_{\mathrm{ex}}-V(t))
    \label{labelIoft}
\end{equation}

Specific values of neuronal parameters used in this work are reported in SM Sec.~\ref{subsec:SM:Parameters}.

\subsection{STDP}
\label{subsec:SM:STDP}

The evolution of plastic synapses implemented in this model is described by spike-timing-dependent plasticity (STDP)~\cite{Morrison_2008, Sboev_2016}, characterized by the pair-based update law:

\begin{equation}
    \Delta w = 
    \begin{cases}
        -W_{-} \cdot (\frac{w}{w_{\mathrm{max}}})^{\mu_{-}} \cdot \exp {\left(-\frac{t_{\mathrm{pre}}-t_{\mathrm{post}}}{\tau_{-}}\right)}\,, & \quad \text{if } t_{\mathrm{pre}}-t_{\mathrm{post}} > 0\\
        W_{+} \cdot (1 - \frac{w}{w_{\mathrm{max}}})^{\mu_{+}} \cdot \exp {\left(-\frac{t_{\mathrm{pre}}-t_{\mathrm{post}}}{\tau_{+}}\right)}\,, & \quad \text{otherwise}
    \end{cases}
\end{equation}

\noindent
where $w_{\mathrm{max}}$ represents a limiting value for the weight and $W_{+}$, $W_{-}$ are constants that set the intensity of weights modification.
The exponents $\mu_{+}$, $\mu_{-}$ can vary in the range $[0,1]$. In the two extreme cases, $\mu_{+/-}=0$ and $\mu_{+/-}=1$, the model is called additive STDP and multiplicative STDP, respectively. Throughout this work, we used a multiplicative STDP rule.

For the purpose of this paper it's convenient to reformulate as below:

\begin{equation}
    \Delta w = 
    \begin{cases}
        -\alpha \lambda \cdot w \cdot \exp {\left(-\frac{t_{\mathrm{pre}}-t_{\mathrm{post}}}{\tau_{-}}\right)}\,, & \quad \text{if } t_{\mathrm{pre}}-t_{\mathrm{post}} > 0\\
        \lambda \cdot (w_{\mathrm{max}} - w) \cdot \exp {\left(-\frac{t_{\mathrm{pre}}-t_{\mathrm{post}}}{\tau_{+}}\right)}\,, & \quad \text{otherwise}
    \end{cases}
    \label{eq:STDP_alfa_lamda}
\end{equation}

\noindent
where $\lambda$ stands for the learning rate and $\alpha$ depicts the asymmetry parameter of depressing and increasing synaptic weights.
According to Hebb's postulate, since the weights of all thalamo-cortical synapses are plastic, if both the input pattern and the contextual signal are kept active for a sufficiently long time, the weights of synapses connecting active thalamic neurons to active cortical neurons will grow. 

The specific values of synaptic dynamics parameters for the different brain states are reported in SM Sec.~\ref{subsec:SM:Parameters}.

\subsection{Model parameters}
\label{subsec:SM:Parameters}

Table~\ref{tab:parameters} reports numerical values for all the parameters nedeed to simulate the model in order to reproduce the present results.

The first simulation stage is a training session in awake regime. Before the training, synapses are set to a $W_0$ value. During the awake training phase, then, synapses are plastic and follow the dynamics specified in SM Sec.~\ref{subsec:SM:STDP}, with $W_{\mathrm{max}}$, $\lambda_{\mathrm{STDP}}$ and the asymmetry factor $\alpha_{\mathrm{STDP}}$ set at the values reported in the \textit{Network Parameters - Awake} subtable. Values for spike-frequency adaptation parameters $b_{\mathrm{SFA}}$ and $\tau_{\mathrm{SFA}}$ are also reported in the same subtable, for both training and classification stages during the awake regime. Notice that synapses that are not plastic in the model remain at the $W_0$ initial value specified in the model during awake training and classification.

External signal is provided to both thalamic and cortical layers in the network though the simulation of Poissonian spike trains (subtable \textit{External stimuli parameters}). A set of independent spike generators characterized by a rate parameter is connected one-to-one to neurons in the network through a static synapse of fixed weight $W$. Specifically, the perceptual signal is provided to the thalamic neurons during both training and classification phases, providing the input encoded according to the pre-processing protocol described in Methods Sec.~\ref{subsec:methods:preprocessing}. During the training phase, an example-specific contextual signal is provided to a subset of cortical excitatory neurons and to all the inhibitory 
cortical neurons (see Methods Sec.~\ref{subsec:methods:combinationContextPeception}). During NREM and REM phases, on the other hand, a non-specific signal (see Methods Sec.~\ref{subsec:methods:brain_states}) is provided to cortical excitatory neurons, whereas no perceptual signal stimulates the thalamic layer. Moreover, during the REM stage, two stimuli are provided to cortical neurons, representing high- and low-frequency stimulations, respectively.

Concerning the neural model parameters, in subtable \textit{Neuron Parameters} we report the capacitance of the membrane $C_{m}$, the slope factor $\Delta_{T}$, the resting membrane potential $E_{L}$, the excitatory and inhibitory reversal potential $E_{\mathrm{ex}}$ and $E_{\mathrm{in}}$, and the leak conductance $g_{L}$. For what concerns time constants, we specify the duration of refractory period $t_{\mathrm{ref}}$ and the excitatory and inhibitory synaptic time constants $\tau_{\mathrm{syn},\mathrm{ex}}$ and $\tau_{\mathrm{syn},\mathrm{in}}$, respectively. Also, the peak potential $V_{\mathrm{peak}}$, the reset potential $V_{\mathrm{reset}}$, and the threshold potential $V_{\mathrm{th}}$ are specified. The other parameters that are not explicitly reported here, are set to the default value of NEST neural model \texttt{aeif}-\texttt{cond}-\texttt{alpha}.

In the \textit{Neuromodulation - Sleep} subtable, we report parameters for transitioning from awake to NREM and REM states. Neuromodulation of excitatory cortico-cortical synapses is achieved by multiplying/dividing (to enter/exit the brain state, respectively) the synaptic distribution by a constant factor, resulting in a scaling down of the average value of synapses encoding for a specific training example to the value shown in the table (See Methods Sec.~\ref{subsec:methods:brain_states} for an explanation of the mechanism). $\left \langle W \right \rangle$ represents the mean value of intra-group synapses after neuromodulation,  with ``intra'' and ``inter'' indicating whether the connection is intra- or inter-area. $\lambda$, $\alpha$ and $\tau_{\mathrm{STDP}}$ refer to the STDP learning rate, asymmetry factor and time scale, respectively (see SM Sec.~\ref{subsec:SM:STDP}, Eq.~(\ref{eq:STDP_alfa_lamda})). $b_{\mathrm{SFA}}$ and $\tau_{\mathrm{SFA}}$ represent the adaptation current and time scale, respectively (see SM Sec.~\ref{subsec:SM:adexNeuron}).
 
Finally, as described in Methods Sec.~\ref{subsec:methods:combinationContextPeception}, the network is trained on examples extracted from the MNIST database of handwritten digit and then exposed to other examples during a classification task. In subtable \textit{MNIST handwritten digits training/classification} we report the number of training and classification examples exposed to the network and the time of exposure. Also, we report the number of classes represented in the dataset and the number of neurons per area trained over the same example.

\begin{table}[!h]

    \caption{
        \textbf{Model Parameters.}
        Values of the parameters used in our simulations. Parameters that are not explicitly reported here, are set to the default value of NEST neural model \texttt{aeif}-\texttt{cond}-\texttt{alpha}. N/A: Not applicable parameters.
    }

    \begin{center}

    \begin{tabular}[t]{|c|c|c|c|}
    \firsthline
    
    \multicolumn{4}{|c|}{\textbf{Network Parameters - Awake}}\\ 
    \hline\hline
    
    \multicolumn{1}{|c|}{\textbf{Connection}} & $\mathbf{W}_0$ & $\mathbf{W}_{\mathrm{max}}$  & $\boldsymbol{\lambda}_{\mathrm{STDP}}$ \\
    \hline
    
    $th \rightarrow cx$ & 0.3 & 3.7 & 0.06\\
    \hline
    $cx \rightarrow th$ & 0.1 & 2 & 0.06\\
    \hline
    $cx ; \ exc \rightarrow exc$ & 0.01 & 75 & 0.06\\
    \hline
    $th ; \ exc \rightarrow inh$ & 3 & N/A & N/A \\
    \hline
    $th ; \ inh \rightarrow exc$ & -1 & N/A & N/A \\
    \hline
    $cx ; \ exc \rightarrow inh_f$ & 200 & N/A & N/A \\
    \hline
    $cx ; \ inh_f \rightarrow exc$ & -20 & N/A & N/A \\
    \hline
    $cx ; \ exc \rightarrow inh_s$ & 3 & N/A & N/A \\
    \hline
    $cx ; \ inh_s \rightarrow exc$ & -20 & N/A & N/A \\
    \hline\hline

    \multicolumn{3}{|l|}{$b_{\mathrm{SFA}}$ (pA)} & 110\\
    \hline

    \multicolumn{3}{|l|}{$\tau_{\mathrm{SFA}}$ (ms)} & 144\\
    \hline

    \multicolumn{3}{|l|}{$\alpha_{\mathrm{STDP}}$ (STDP asymmetry factor)} & 1\\
    \hline\hline
    
    \hline\hline
    
    \multicolumn{4}{|c|}{\textbf{Neuromodulation - Sleep}}\\ 
    \hline\hline

    \multicolumn{2}{|c|}{\textbf{Parameter}} & \textbf{NREM} & \textbf{REM} \\
    \hline
    \multicolumn{2}{|c|}{$\left \langle W \right \rangle ; \ th \rightarrow cx \; (intra)$} & $50$ & $50$\\
    \hline
    \multicolumn{2}{|c|}{$\left \langle W \right \rangle ; \ cx \rightarrow th \; (intra)$} & $0.186$ & $1.488$\\
    \hline
    \multicolumn{2}{|c|}{$\left \langle W \right \rangle_{\mathrm{CA}} ; \ cx \rightarrow cx \; (intra)$} & $59$ & $as \ awake$\\
    \hline
    \multicolumn{2}{|c|}{$\left \langle W \right \rangle_{\mathrm{CA}} ; \ cx \rightarrow cx \; (inter)$} & $0$ & $97.5$\\
    \hline
    \multicolumn{2}{|c|}{$cx ; \ inh_f \rightarrow exc$} & $-0.5$ & $-0.5$\\
    \hline
    \multicolumn{2}{|c|}{$cx ; \ inh_s \rightarrow exc$} & $-0.5$ & $-400$\\
    \hline
    \multicolumn{2}{|c|}{$\lambda_{\mathrm{STDP}} \; ; \ cx \rightarrow cx$} & $3\cdot 10^{-4}$ & $3\cdot 10^{-4}$ \\
    \hline
    \multicolumn{2}{|c|}{$\alpha_{\mathrm{STDP}} \; ; \ cx \rightarrow cx$} & $3$ & $3$\\
    \hline
    \multicolumn{2}{|c|}{$\tau_{\mathrm{STDP}}$ (ms)} & 400 & 400 \\
    \hline
    \multicolumn{2}{|c|}{$b_{\mathrm{SFA}}$ (pA)} & 200 & 50 \\
    \hline
    \multicolumn{2}{|c|}{$\tau_{\mathrm{SFA}}$ (ms)} & 400 & 400 \\
    \hline
    
    \end{tabular}
    \begin{tabular}[t]{|c|c|c|c|}
    \firsthline
    
    \multicolumn{4}{|c|}{\textbf{External stimuli parameters}}\\ 
    \hline\hline
    
    \multicolumn{1}{|c|}{\textbf{Connection}} & {\textbf{State}} & {\textbf{Rate} Hz} & $\mathbf{W}$\\
    \hline

    $\rightarrow cx ; \ exc$ & training &  600 & 500\\
    \hline
    $\rightarrow cx ; \ inh$ & training &  80 & 320\\
    \hline
    $\rightarrow th ; \ exc$ & training &  40 & 400\\
    \hline
    $\rightarrow th ; \ exc$ & classification &  40 & 400\\
    \hline
    $\rightarrow cx ; \ exc$ & NREM &  1000 & 13.3\\
    \hline
    $\rightarrow cx ; \ exc$ & REM &  7.2 & 130\\
    \hline
    $\rightarrow cx ; \ exc$ & REM &  30 & 60\\
    \hline\hline

    \multicolumn{4}{|c|}{\textbf{Neuron Parameters}}\\ 
    \hline\hline

    \multicolumn{3}{|l|}{$C_m$ (pF)} & 281 \\
    \hline    
    \multicolumn{3}{|l|}{$\Delta_T$ (mV)} & 2 \\
    \hline
    \multicolumn{3}{|l|}{$E_L$ (mV)} & -70.6 \\
    \hline
    \multicolumn{3}{|l|}{$E_{\mathrm{ex}}$ (mV)} & 0 \\
    \hline
    \multicolumn{3}{|l|}{$E_{\mathrm{in}}$ (mV)} & -85 \\
    \hline
    \multicolumn{3}{|l|}{$g_L$ (ns)} & 30 \\
    \hline
    \multicolumn{3}{|l|}{$t_{\mathrm{ref}}$ (ms)} & 2 \\
    \hline
    \multicolumn{3}{|l|}{$\tau_{\mathrm{syn},\mathrm{ex}}$ (ms)} & 0.2 \\
    \hline
    \multicolumn{3}{|l|}{$\tau_{\mathrm{syn},\mathrm{inh}}$ (ms)} & 2 \\
    \hline
    \multicolumn{3}{|l|}{$V_{\mathrm{peak}}$ (mV)} & 0 \\
    \hline
    \multicolumn{3}{|l|}{$V_{\mathrm{reset}}$ (mV)} & -60 \\
    \hline
    \multicolumn{3}{|l|}{$V_{\mathrm{th}}$ (mV)} & -50.4 \\
    \hline
    \hline
    
    \multicolumn{4}{|c|}{\textbf{MNIST handwritten digits}}\\
    \multicolumn{4}{|c|}{\textbf{training/classification}}\\
    \hline
    \hline

    \multicolumn{3}{|l|}{\# of digit classes} & 10 \\
    \hline    
    \multicolumn{3}{|l|}{\# of training examples per class} & 5 \\
    \hline
    \multicolumn{3}{|l|}{\# of neurons in example-specific group} & 20 \\
    \hline
    \multicolumn{3}{|l|}{\# of test examples} & 210 \\
    \hline
    \multicolumn{3}{|l|}{Presentation time during training (s)} & 1.5 \\
    \hline
    \multicolumn{3}{|l|}{Presentation time during test (s)} & 1 \\
    \hline
    \multicolumn{3}{|l|}{Training/test rest time (s)} & 1.5 \\
    \hline
    
    \end{tabular}
    \label{tab:parameters}

    \end{center}
\end{table}

\end{document}